\documentclass[runningheads]{aa}  

\usepackage{graphicx}
\usepackage{txfonts}
\usepackage{ulem}
\usepackage[pdfencoding=auto,psdextra]{hyperref}
\hypersetup{
    colorlinks=true,
    linkcolor=blue,
    filecolor=magenta,      
    urlcolor=blue,
    citecolor=blue
}

\usepackage{xcolor}
\usepackage{lipsum}
\usepackage{todonotes}
\usepackage{xspace}

\newcommand{\Mpch}{\ensuremath{h^{-1}\; \text{Mpc}}\xspace}

\begin{document}

    \title{A Bayesian catalog of 100 high-significance voids in the Local Universe}

   \author{R. Malandrino
          \inst{1}\fnmsep\thanks{Corresponding author, \email{rosa.malandrino@iap.fr}},
          G. Lavaux
          \inst{1},
          B. D. Wandelt\inst{1, 2,3,4},
          S. McAlpine\inst{5},
          \and
          J. Jasche \inst{5}
          }
\authorrunning{Malandrino et al.}
\titlerunning{A Bayesian catalog of high-significance voids}

   \institute{Sorbonne Université, CNRS, UMR 7095, Institut d'Astrophysique de Paris, 98bis Bd Arago, 75014, Paris, France
         \and
             Department of Physics and Astronomy, Johns Hopkins University, 3400 North Charles Street, Baltimore, MD 21218, USA
        \and
             Department of Applied Mathematics and Statistics, Johns Hopkins University, 3400 North Charles Street, Baltimore, MD 21218, USA
        \and
            Center for Computational Astrophysics, Flatiron Institute, 162 5th Avenue, New York, NY 10010, USA
        \and
            The Oskar Klein Centre, Department of Physics, Stockholm University, Albanova University Center, 106 91 Stockholm, Sweden
            }

   \date{Received XXXX; accepted YYYY}

 
  \abstract
   { 
   While cosmic voids are now recognized as a valuable cosmological probe, identifying them in a galaxy catalog is challenging for multiple reasons: observational effects such as holes in the mask or magnitude selection hinder the detection process; galaxies are biased tracers of the underlying dark matter distribution; and it is non-trivial to estimate the detection significance and parameter uncertainties for individual voids.

   }
   { Our goal is to extract a catalog of voids from constrained simulations of the large-scale structure that are consistent with the observed galaxy positions, effectively representing statistically independent realizations of the probability distribution of the cosmic web. This allows us to carry out a full Bayesian analysis of the structures emerging in the Universe. }
   { We use 50 posterior realizations of the large-scale structure in the \texttt{Manticore-Local} suite, obtained from the 2M++ galaxies, with $z \lesssim 0.1$ and a sky area between $\sim 23 \ 000 - 36 \ 000 \ \text{deg}^2$.
   Running the \texttt{VIDE} void finder on each realization, we extract 50 independent void catalogs. 
   We perform a posterior clustering analysis to identify high-significance voids at the $5\sigma$ level, and we assess the probability distribution of their properties combining the contributions of independent large-scale structure realizations.
   }
   { We produce a catalog of 100 voids with high statistical significance, available at \url{https://voids.cosmictwin.org/}, including the probability distributions of the centers and radii of the voids. We characterize the morphology of these regions, effectively producing a template for density environments that can be used in astrophysical applications such as galaxy evolution studies. }
   { While providing the community with a detailed catalog of voids in the nearby Universe, this work also constitutes an approach to identifying cosmic voids from galaxy surveys that allows us to account rigorously for the observational systematics intrinsic to direct detection, and provide a Bayesian characterization of their properties. }

   \keywords{Cosmology: large-scale structure --
                Methods: statistical --
                Methods: numerical --
                Catalogs
               }

    \maketitle

%

\begin{figure*}
    \centering
    \includegraphics[width=\textwidth]{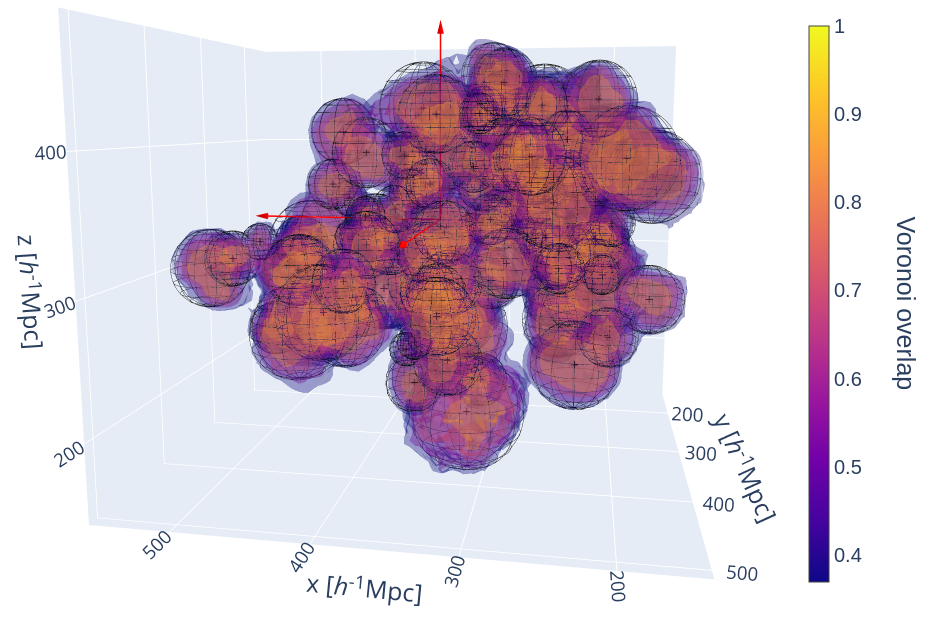}
   
    \caption{
        Visual representation of our catalog of 100 voids in the Local Neighborhood within a distance of $\approx 200 \Mpch$. The colorscale represents the Voronoi overlap rate, i.e. how often a point is contained in a void throughout different realizations. This quantity is loosely related to the density profile of voids, with high values representing the innermost underdense regions, while lower values are at the edges. The clouds are truncated at the value of $0.37$ in order to preserve the statistical volume of all voids. Finally, the wireframe spheres mark the effective volume of each void. The cartesian axis are centered in the observer, with the $xy$ plane corresponds to the equatorial plane, and the $\hat{z}$ axis pointing to the equatorial North Pole. At very nearby distances the method is able to reconstruct voids lying behind the galactic plane, a region not observed in the survey.
        An interactive version of this plot can be found here: \url{https://voids.cosmictwin.org}
        }
    \label{fig:allVoids}
\end{figure*}

\section{Introduction}
\label{sec:intro}

Cosmic voids are the largest objects emerging in the cosmic web, covering the majority of the volume of the Universe.
They present fundamentally different features compared to overdensities of the dark matter distribution, such as clusters, walls, and filaments, thus representing a unique environment to probe different aspects of astrophysics and cosmology.

In the past decade, voids have become a well-established probe to gather cosmological information from the large-scale structure, complementing the constraints inferred from traditional clustering analysis \citep{Pisani2019, Hamaus2020, Kreisch2022, Moresco2022}.
As their matter content is low by definition, the evolution of voids is dominated by the mysterious cosmic acceleration, making them a pristine environment to constrain dark energy physics \citep{Lee2009, Biswas2010,Lavaux2010, Bos2012, Pollina2015}, or test theories of gravity at the largest scales \citep{Hui2009, Clampitt2013, Cai2015, Zivick2015, Hamaus2016, Falck2017, Contarini2021}.
Similarly, the gravitational effect of neutrinos is not negligible in voids, and the properties of these structures depend on the mass of this elusive particle \citep{Massara2015, Kreisch2019, Schuster2019, Zhang2020, Bayer2021}.
The most powerful tools to estimate cosmological parameters are voids summary statistics such as their size distribution \citep{Pisani2015, Contarini2019, Verza2019, Contarini2022, Contarini2023}, the analysis of the void-galaxy correlation, equivalent to their density profiles \citep{Hamaus2014b, Hamaus2014a}, and the Alcock-Paczy\'nski test \citep{Alcock1979, Lavaux2012, 10.1007/978-1-4614-3520-4_3, Sutter2012, Sutter2014AP, Hamaus2015, Hamaus2022}.
Additionally, voids are a relatively quiet environment in the cosmic landscape, unaffected by the complex dynamical interaction that characterizes clusters, making them relatively easy to model \citep{VANDEWEYGAERT2011, FShandarin2011}. As a consequence, the primordial features imprinted in the density perturbations by inflation remain almost untouched in voids throughout cosmic history: probing the substructure of underdense regions of space yields information on the early Universe, namely deviations from a perfectly Gaussian distribution of the initial conditions \citep{Kamionkowski2009, Lam2009, DAmico2011, Uhlemann2017, Chan2019}.
Finally, as part of the large-scale structure, voids have an impact on the path of photons emitted from background sources. 
An additional source of cosmological information lies in cross-correlations of their features with the weak gravitational lensing of galaxies \citep{Krause2012, Melchior2014, Barreira2015, Sanchez2016, Davies2021}, and with the Cosmic Microwave Background secondary anisotropies, namely the Integrated Sachs-Wolfe (ISW) effect \citep{Inoue2007, Granett2008, Ili2013, Cai2014, Kovacs2019}, the thermal and kinematic Sunyaev-Zel'dovich (SZ) effect \citep{GarciaBellido2008, Ichiki2016, Hoscheit2018}, and density maps reconstructed from cosmic shear \citep{PlanckLensing, Vielzeuf2020, Sartori2024}.

\bigskip

Beyond cosmology, voids are interesting laboratories for numerous astrophysical applications.
For instance, galaxies populating voids present different properties than those found in overdensities, such as luminosity function \citep{Rojas2004, Hoyle2005, Hoyle2012, Moorman2015} and star formation activity \citep{Ricciardelli2014, Moorman2016, Beygu2016, DomnguezGmez2023}.
As halo and void galaxies undergo different formation histories, differentiating between the two environments can shed light on galaxy evolution \citep{Kreckel2010, VANDEWEYGAERT2011, Kraljic2017, Martizzi2019, Lazar2023,RodrguezMedrano2024, Pan2025}.
Additionally, the relationship between galaxies and the underlying matter distribution, often referred to as galaxy bias, is not fully understood \citep{Wechsler2018}, and can be further investigated in different clustering regimes, including voids \citep{Pollina2017, Pollina2019}.
As an example, dwarf galaxies are known to be more evenly distributed than massive galaxies, making voids the perfect environment to study this class of galaxies \citep{Dekel1986, Eder1989, Shull1996, Karachentseva1999, Hoeft2006, McConnachie2012}.

The occurrence of supernovae correlates with the cosmic web \citep{Tsaprazi2021}, with recent studies focusing on these events exclusively inside voids \citep{Aubert2025}. Supernovae are a standard probe to estimate the Hubble constant in the Local Universe, which is in tension with early time measurements \citep{Verde2019, DiValentino2021}.
A violation of the Copernican principle, positioning the Milky Way in a particularly underdense region of the Universe, has been proposed to explain the apparent faster expansion of the Local Universe \citep{Wojtak2013, Sundell2015, Ding2020, Haslbauer2020}. It is still a matter of debate how relevant this is for the Hubble tension  \citep{Kenworthy2019, Cai2021, Camarena2022, Castello2022}, thus a better understanding of neighboring voids could further test this hypothesis.

Voids also have the potential to probe high-energy astrophysics, namely by studying the properties of black holes and Active Galactic Nuclei (AGN) residing in void galaxies \citep{Porqueres2018, Habouzit2020, Ceccarelli2021, Mishra2021, Oei2024}, and the electron-positron pair beam produced by blazars \citep{Schlickeiser2012a, Schlickeiser2012b, Miniati2013};
these objects interact with the magnetic fields that have been detected throughout the cosmic web \citep{Neronov2010, Tavecchio2010, Taylor2011}:
analyzing them in cosmic voids can help test the hypothesis of a primordial origin and put constraints on magnetogenesis mechanisms \citep{Banerjee2004, Subramanian2016, Hutschenreuter2018, Korochkin2022, Hosking2023}.

\bigskip

Despite the enormous value of cosmic voids for astrophysical and cosmological applications, observational and theoretical issues have prevented the community from fully exploiting their potential.
Ever since the first discovery by \citet{Gregory1978}, and the pioneering works from \citet{Joeveer1978, Kirshner1981, deLapparent1986, TullyFisher}, astronomers have looked into the distribution of galaxies in redshift surveys to identify these structures.
Since then, several catalogs with hundreds or thousands of statistical voids have been published (e.g. \citealt{Pan2012, Sutter2012catalog, Nadathur2014, Sutter2014catalog, Nadathur2016, Mao2017, Aubert2022, Douglass2023}), yet only a handful of voids are truly agreed upon and referred to in the literature with proper names (e.g. the Local Void).

The field has notably grown in the past decades, but finding voids and characterizing their features remains non-trivial due to intrinsic challenges.
First, we can only identify voids in the cosmic web through biased tracers of the underlying dark matter distribution, such as galaxies: the central underdensities of voids in the two fields do not strictly match, but can be mapped into one another with a parametric transformation \citep{Hamaus2014b, Sutter2014}.
Unfortunately, redshift surveys are affected by observational issues, such as holes in the mask and magnitude limits.
The galaxies occupying dark matter overdensities at the edge of a particular survey might be too faint to be detected by the instrument, and the apparent emptiness of the corresponding region of space could be mistaken for a void.
Conversely, excluding unobserved regions from the void finding procedure has an effect on the resulting voids and their statistical properties, depending on the sparsity of the samples and the irregularities of the mask \citep{Sutter2014catalog}.

A certain degree of ambiguity would remain even with perfect knowledge of the galaxy positions. Voids are not discrete and localized structures like halos; rather, they are extended and interconnected regions that spill into each other through empty tunnels.
Multiple definitions and void finding algorithms exist in the literature (e.g. \citealt{Hoyle2004, Padilla2005, Platen2007, zobov, Lavaux2010, Cautun2012, vide}), and the respective voids will inevitably present different features \citep{Colberg2008}.
Even the few famous voids such as the Local Void do not have well-defined centers and boundaries, but a very complex morphology \citep{Tully2019}.
Finally, direct observations deal with a single Universe, rather than multiple realizations of the underlying statistical process generating the large-scale structure, making the significance of the void detection and parameter uncertainties non-trivial to estimate.

\bigskip

In this work, we present a novel statistical method to define a catalog of high-significance voids working around the aforementioned observational issues.
We employ state-of-the-art constrained simulations of the Universe from the \texttt{Manticore Local} \citep{manticore} posterior realizations of the cosmic web: these simulations are the most precise reconstruction to date of our Local Neighborhood, making the voids extracted from it likely to be real in the dark matter distribution, as opposed to being an artifact of observational surveys.
Other works using the initial conditions of constrained simulations to extract void catalogs can be found in \citet{Leclercq2015, Desmond2022, Stopyra2024}.
The framework is fully Bayesian, relying on the \texttt{BORG} algorithm \citep{borg, Jasche2015, Lavaux2015, Lavaux2019} for the inference, and as such we are able to assess the significance of the detected structures, and to characterize the statistical uncertainties of our results. 
The final catalog consists of well-defined regions of space, with identifiable centers and boundaries, making it precise and easy to use for a reliable characterization of the density environment. A visualization is presented in Figure~\ref{fig:allVoids}, with the interactive version presented on our dedicated website\footnote{\url{https://voids.cosmictwin.org/}}. The catalog is available publicly for download\footnote{\url{https://github.com/RosaMalandrino/LocalVoids/}}.

In Section~\ref{sec:data} we present the constrained simulations used as posterior samples of the Local Universe, and the algorithm to extract voids from them.
In Section~\ref{sec:clustering} we describe the statistical methodology to detect and characterize high-significance voids, while in Section~\ref{sec:analysis} we discuss the quality of the produced catalog.
In Section~\ref{sec:comparison} we compare with known voids in the literature, while in Section~\ref{sec:conclusion} we discuss the whole work and present our conclusions.


\begin{figure*}
    \begin{minipage}[t]{0.45\linewidth}
        \flushleft
        \includegraphics[height=8.3cm]{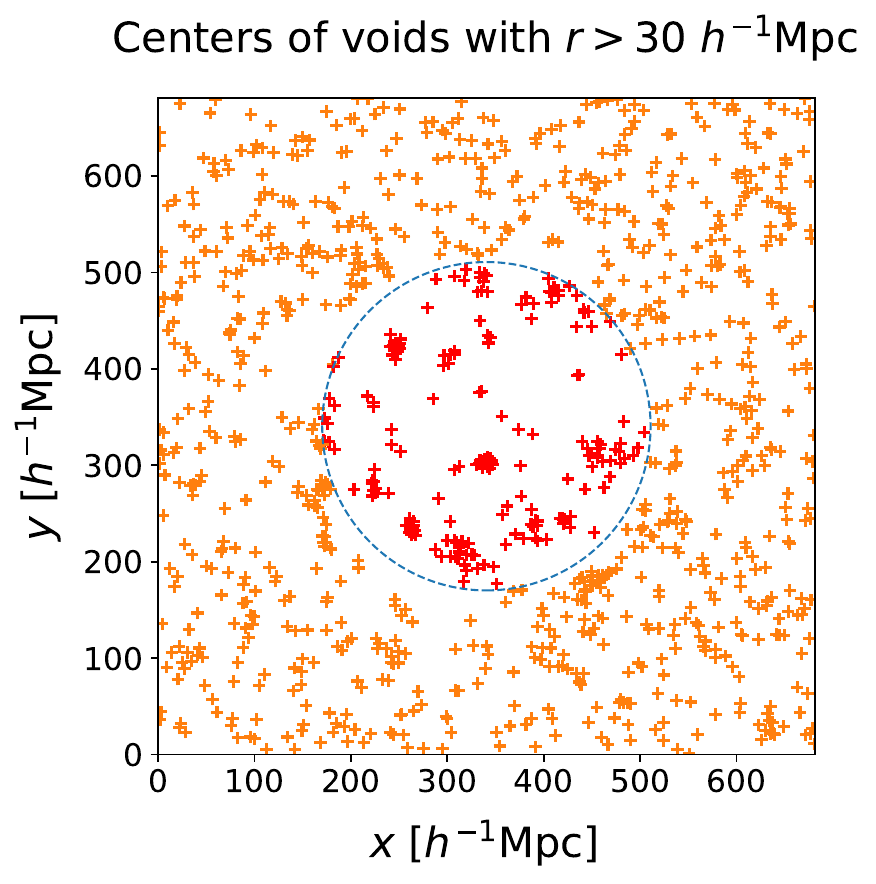}
    \end{minipage}
    \hspace{0.5cm}
    \begin{minipage}[t]{0.55\linewidth}
        \flushleft
        \includegraphics[height=8.22cm]{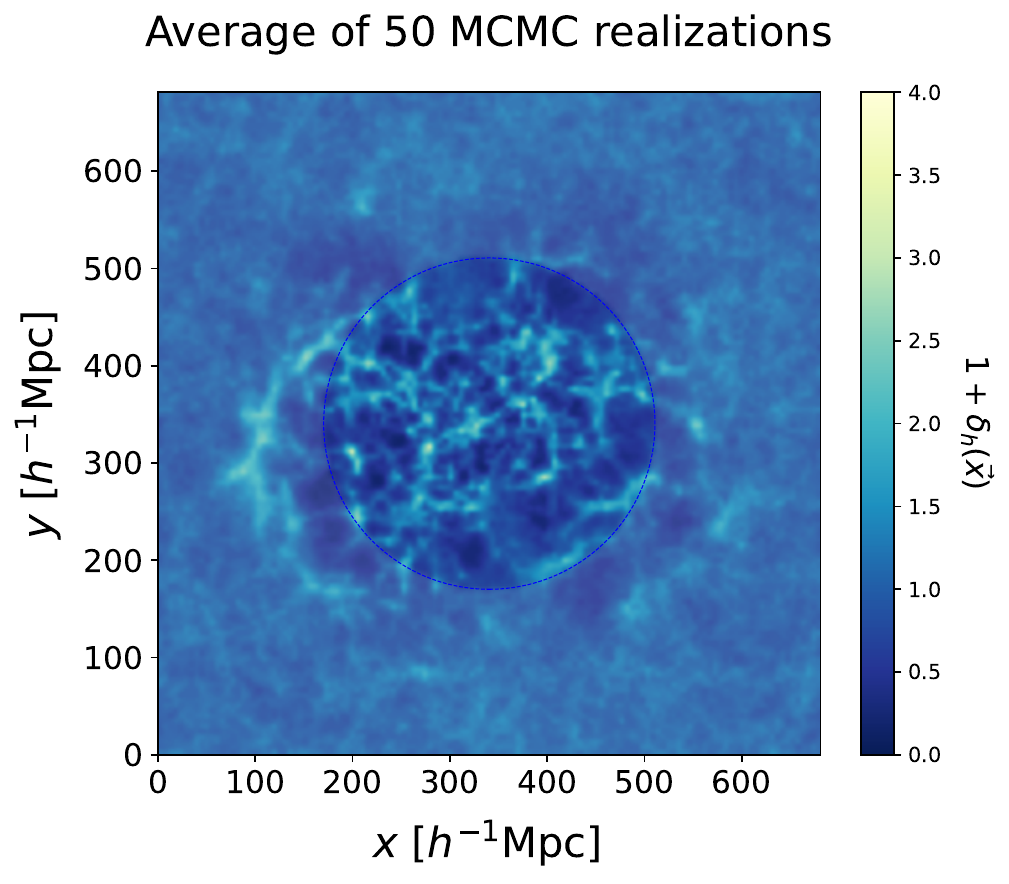}
    \end{minipage}
    \caption{\\
    Left panel: centers of voids with radius $r > 30\,\Mpch$ overlapped on a slice of the 3D space. The dashed blue circle corresponds to a radius of $\approx 250 \, \text{Mpc}$, encompassing the volume with observed galaxies - where the constraints are more reliable - with interior points marked in red, the others in orange. The figure shows that independent posterior realizations of the large-scale structure present comparable features, with similarly sized voids occupying the same regions of space. The detection of stable voids between Markov Chain Monte Carlo (MCMC) realizations reduces to a clustering problem. The unconstrained region outside the circle present a random distribution and can be used to assess the probability of having a cluster of $n$ points by random chance. \\
    Right panel: halo field averaged over all the MCMC realizations. The underdense regions visibly match the clusters in the overlapped centers.
              }

    \label{fig:centersOverlap}
\end{figure*}

\section{Data products}
\label{sec:data}

To characterize the voids in our Local Neighborhood, we use a set of constrained N-body simulations of the large-scale structure, initialized to be consistent with the galaxy positions identified in the 2M++ data compilation \citep{2Mpp}.
Produced in the Bayesian framework of the \texttt{BORG} algorithm, they represent 50 independent samples of the posterior distribution of the large-scale structure and allow us to carry out a complete statistical analysis of the voids. These constrained simulations will be made publicly available\footnote{\url{https://cosmictwin.org}}.

\subsection{Posterior realizations of the large-scale structure}
\label{subsec:manticore}

To describe the Local Universe in a Bayesian framework, we use the \texttt{Manticore} Local posterior simulations \citep{manticore}, a set of posterior realizations of the large-scale structure in our cosmic neighborhood.
They are produced following two separate procedures.

The first step consists of a Bayesian inference within the \texttt{BORG} algorithm.
This probabilistic engine samples initial conditions of the Universe, evolves them to the present day distribution and constrains the realizations that are consistent with observed galaxy counts and positions.
The initial power spectrum is computed with \texttt{CLASS} \citep{CLASS}, with the cosmological parameters inferred from the Dark Energy Survey year three \citep{DES} '3x2pt + All Ext.' $\Lambda$CDM cosmology: $h = 0.681$, $\Omega_m = 0.306$ , $\Omega_\Lambda = 0.694$, $\Omega_b = 0.0486$, $A_s = 2.099\times10^{-9}$, $n_s = 0.967$, $\sigma_8 = 0.807$.

We use a set of 50 independent posterior samples of the initial conditions (IC) that might have seeded the galaxy distribution in our local neighborhood at present day.
They are represented on a box with side $L = 1~\text{Gpc} = 681\,\Mpch$, on a grid of $N = 256^3$ voxels.
The grid is then augmented to $N = 1024^3$ particles with mass resolution of $3.7 \times 10^{10} \, h^{-1} \ M_{\odot}$. The IC are generated with second order Lagrangian Perturbation Theory (2LPT), and their evolution to $z = 0$ is computed with the \texttt{SWIFT} N-body solver \citep{SWIFT}.
Finally, structures are extracted from the present-day dark matter distribution using the \texttt{HBT+} (Hierarchical Bound Tracing) subhalo finder \citep{Han2012, Han2018}. 
The mass of the halos is defined as $M_{200, \text{crit}}$, i.e. the mass enclosed within $r_{200, \text{crit}}$, the radius within which the mean enclosed density is 200 times the critical density of the Universe ($200 \times \rho_\text{crit}$).
The result is 50 independent distributions, consisting of $\sim 1.3 \times 10^6$ halos with resolution $M_\text{min} = 10^{12} M_{\odot}  $, that can be then used as tracers to identify voids. We refer to the full Manticore paper \citep{manticore} for further details regarding the construction of the constrained halo catalogs.

\begin{table*}

\caption{
            Summary of void properties for a few example voids. The first 4 columns represent the coordinates of the center in the sky, in terms of equatorial coordinates and distance.
            The fourth column mentions the name of the matches we find in the literature, often corresponding to the constellation located in the direction of the void.
            NLV indicates the Northern Local Void of \citet{Einasto1994}.
            A more extended discussion of this comparison is presented in Section~\ref{sec:comparison}. 
            Column 5 represents the redshift range from the nearest to the farther edge of the void along the line of sight, assuming the effective radius as isotropic size.
            Columns 6-9 represent summary properties of the posterior distributions for the effective radius and the center position, using the mean and standard deviation.
            In our cartesian coordinates system the observer is located at $[340.5, 340.5, 340.5]$ \Mpch, with the $xy$ plane corresponding to the equatorial plane, and the $\hat{z}$ axis pointing to the equatorial North Pole.
            The full table with all 100 voids of the catalog can be found in Appendix~\ref{app:catalog}.    
            }
\label{tab:voids_catalog_short}    
\centering          
\begin{tabular}{c | c c c c| c | c | c c c}     
\hline\hline 
 & \multicolumn{4}{c | }{equatorial sky coordinates} & \multicolumn{1}{c | }{redshift }  & \multicolumn{1}{c | }{radius $r$} & \multicolumn{3}{c}{position $\vec{x}$}\\ 
\hline
Void & $\alpha$ &  $\delta$ &  $d_c$ & Void & $z$ range &  $\bar{r}$ & $\bar{x}$ & $\bar{y}$ & $\bar{z}$\\

ID & $[\mathrm{hms}]$ &  $[^\circ]$ &  $[\Mpch ]$ & Match &  $[ \text{km s}^{-1} ]$ &  $[ \Mpch ]$ & $[ \Mpch ]$ & $[\Mpch ]$ & $[ \Mpch ]$\\

\hline

    $ 10 $ & $ 18^\text{h}05^\text{m}$ & $ -16.2$ & $ 38.8$ & Local & $ 500 - 7 \, 300 $ & $33.9 \pm 3.0$ & $341.3 \pm 4.9$ & $303.2 \pm 3.4$ & $329.7 \pm 3.5$ \\
    $ 75 $ & $ 00^\text{h}52^\text{m}$ & $ 43.9$ & $ 33.0$ & Local & $ 1 \, 500 - 5 \, 100 $ & $18.0 \pm 2.0$ & $363.7 \pm 3.6$ & $345.8 \pm 2.1$ & $363.4 \pm 3.7$ \\
    $ 83 $ & $ 17^\text{h}26^\text{m}$ & $ -10.9$ & $ 96.0$ & NLV & $ 8 \, 300 - 11 \, 100 $ & $14.0 \pm 2.6$ & $326.7 \pm 2.6$ & $247.2 \pm 2.3$ & $322.4 \pm 2.5$ \\
    $ 45 $ & $ 12^\text{h}48^\text{m}$ & $ 12.9$ & $ 56.1$ & Coma & $ 3 \, 700 - 7 \, 500 $ & $19.0 \pm 2.8$ & $287.0 \pm 4.1$ & $329.2 \pm 2.6$ & $353.0 \pm 3.1$ \\
    $ 88 $ & $ 14^\text{h}51^\text{m}$ & $ 39.4$ & $ 138.6$ & Boötes & $ 10 \, 000 - 18 \, 100 $ & $39.5 \pm 2.9$ & $261.9 \pm 10.7$ & $267.7 \pm 10.0$ & $428.5 \pm 9.3$ \\
    $ 15 $ & $ 23^\text{h}48^\text{m}$ & $ 16.6$ & $ 95.3$ & Pisces & $ 7 \, 500 - 11 \, 700 $ & $20.9 \pm 3.1$ & $431.8 \pm 4.7$ & $335.6 \pm 5.6$ & $367.7 \pm 2.9$ \\
    $ 57 $ & $ 11^\text{h}01^\text{m}$ & $ -29.4$ & $ 65.8$ & Hydra & $ 3 \, 800 - 9 \, 400 $ & $27.9 \pm 3.0$ & $285.1 \pm 4.0$ & $355.2 \pm 6.1$ & $308.2 \pm 2.6$ \\
    $ 28 $ & $ 10^\text{h}59^\text{m}$ & $ 2.8$ & $ 43.6$ & Leo & $ 2 \, 300 - 6 \, 400 $ & $20.2 \pm 2.0$ & $298.4 \pm 4.0$ & $351.9 \pm 4.2$ & $342.6 \pm 3.0$ \\

\hline                  
\end{tabular}
\end{table*}

\subsection{Independent void catalogs}
\label{subsec:vide}

We identify voids in the 50 samples using \texttt{VIDE}\footnote{\url{https://code.cosmicvoids.net}} \citep{vide}, a void-finder which implements an enhanced version of \texttt{ZOBOV} \citep{zobov} to construct voids with a watershed algorithm.

\texttt{VIDE} computes a Voronoi tesselation of the tracers of the density field - in our case halos - and merges the cells into zones and voids using the watershed transform \citep{Platen2007}.
The resulting voids consist of the union of multiple cells, and as such they have complicated shapes, but can be summarized with their centers and effective radii.
The center of a void is computed as the center of its cells, weighted by their volumes:
$$\vec{x}_\text{center} = \frac{1}{\sum_i V_i} \sum_i \vec{x}_i V_i , $$
where $\vec{x}_i$ and $V_i$ are the position and volume of each Voronoi cell.
The effective radius is:
$$ R_\text{eff} = \left(\frac{3}{4\pi} V \right)^\frac{1}{3} , $$
where $V$ is the sum of the volume of all the Voronoi cells.
These two properties are automatically computed by \texttt{VIDE} and printed in the final catalogs.
Moreover, the code saves the volume of the Voronoi cells corresponding to each tracer, which can then be used to reconstruct the actual shape of each void.

We note that there is no unique definition of voids: as a consequence, the properties of these structures inevitably depend on the algorithm chosen to identify them.
\texttt{VIDE} is one of the most established void finders in the literature, its features have been carefully tested (e.g. \citealt{Sutter2014bis, Sutter2014catalog}), and it has been successfully employed in numerous independent works.
It belongs to the family of watershed-based void finders, as described in e.g. \citet{Platen2007, zobov}. This kind of void finders have characteristics that are best suited for our needs, as opposed to other void finders based on spherical underdensities \citep{Hoyle2004, Padilla2005}, or on the dynamics of the tracers of the density fields \citep{Lavaux2010, Cautun2012}.
The former category assumes that voids are spheres, making their identification simpler and faster, at the cost of losing information on their morphology;
the latter reconstructs voids as sources of the velocity field, and as such contains information on the dynamics.
However, obtaining reliable estimates of velocities from observation is a challenging task, while reconstructing them from simulations is computationally expensive.
Additionally, such void finders miss the so-called "voids-in-clouds" \citep{Sheth2004} - i.e. underdense bubbles that are evolving into clustered structure, like walls, filaments or halos - where the velocities point inward to the center of the void, making them appear overdense in the divergence field.
\texttt{VIDE} provides a good trade-off between complexity in void morphology and computational performance: extracting catalogs on the $50$ simulations boxes of \texttt{Manticore-Local} with periodic boundary conditions - with $\sim 1.3$ million halos each - can be achieved in the order of magnitude of a few hours.


\section{Statistical void detection}
\label{sec:clustering}

We compare the 50 independent catalogs obtained from different realizations of the large-scale structure, in order to identify which voids occur multiple times in the same place. 
Two voids belonging to the same cosmic web realization cannot overlap spatially due to the watershed void definition; however, a true void in the Universe will be identified as an underdensity in multiple Markov Chain Monte Carlo (MCMC) samples.
Thus, placing all void centers from independent catalogs on the same shared three-dimensional space will produce clusters of points centered on real voids in the Universe.
This signal can be noisy due to local minima of the density field being identified as voids by \texttt{VIDE} due to the scatter of tracers.
For this reason, we only use the default catalog of \texttt{VIDE}, which excludes shallower voids with higher central density. This choice effectively reduces the risk of counting spurious ones, which they are not well understood \citep{zobov, Cousinou2019}.
We also choose to compare only similarly sized voids, which are more likely to be candidates for the same true void in the Universe. 

The left panel of Fig.~\ref{fig:centersOverlap} shows the positions of the centers of voids with radius $r > 30\,\Mpch$ on a slice of the 3D space, with the blue circle representing the volume occupied by observed galaxies, up to a distance of $\approx 250 \, \text{Mpc}$. For features in the matter distribution that are strongly supported by the data, 
independent posterior realizations of the large-scale structure will show similarities, with voids of comparable sizes occupying the same regions of space.
This is also evident when observing the average density of all realizations, as shown on the right panel of Fig.~\ref{fig:centersOverlap}.
The detection of stable voids between samples of the posterior reduces to a clustering problem.

\begin{figure*}
\begin{minipage}[t]{0.5\linewidth}
        \flushleft
        \includegraphics[width=\linewidth, trim={3cm 3cm 5cm 5cm},clip]{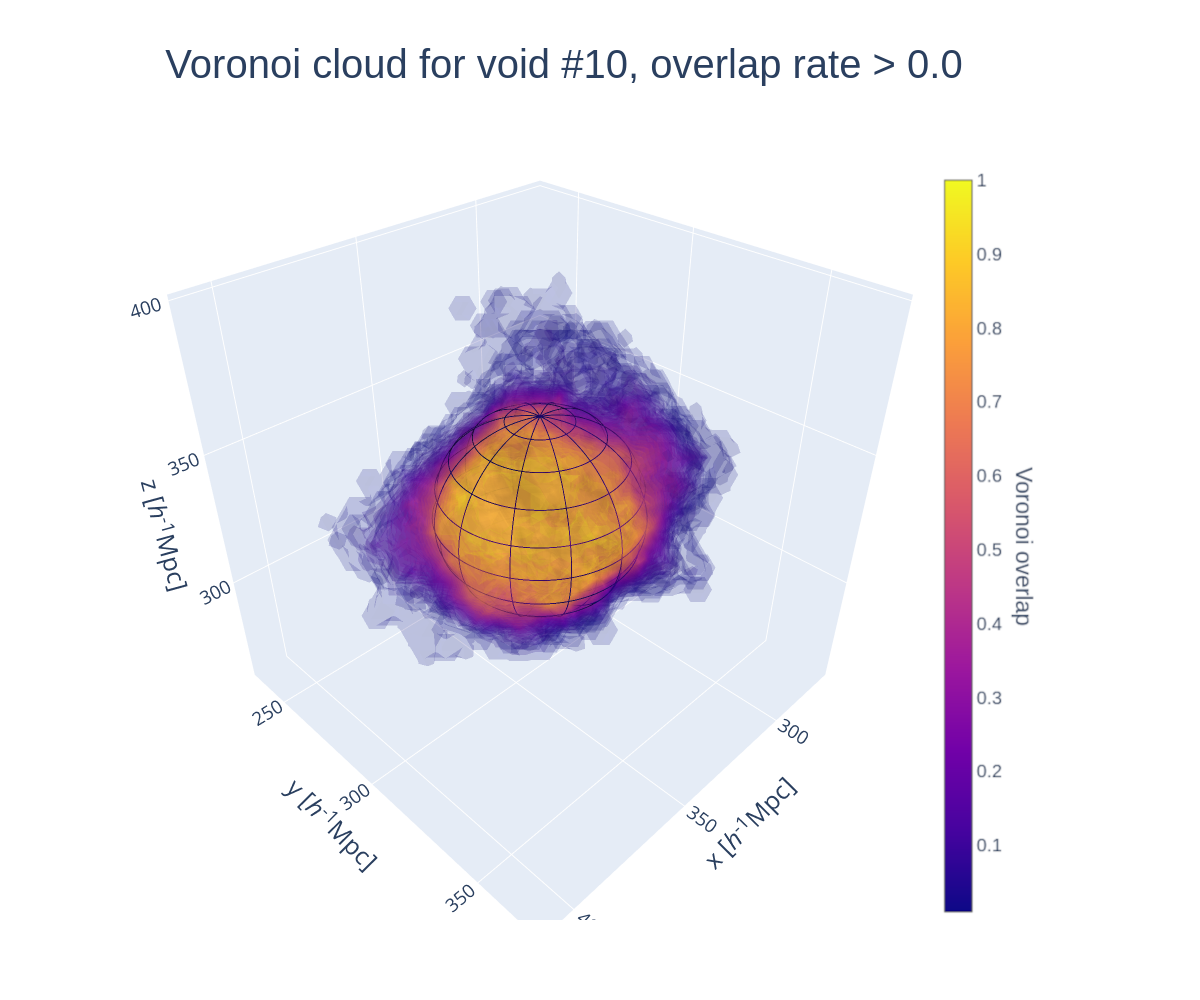}
    \end{minipage}
    \begin{minipage}[t]{0.5\linewidth}
        \flushright
        \includegraphics[width=\linewidth, trim={3cm 3cm 5cm 5cm},clip]{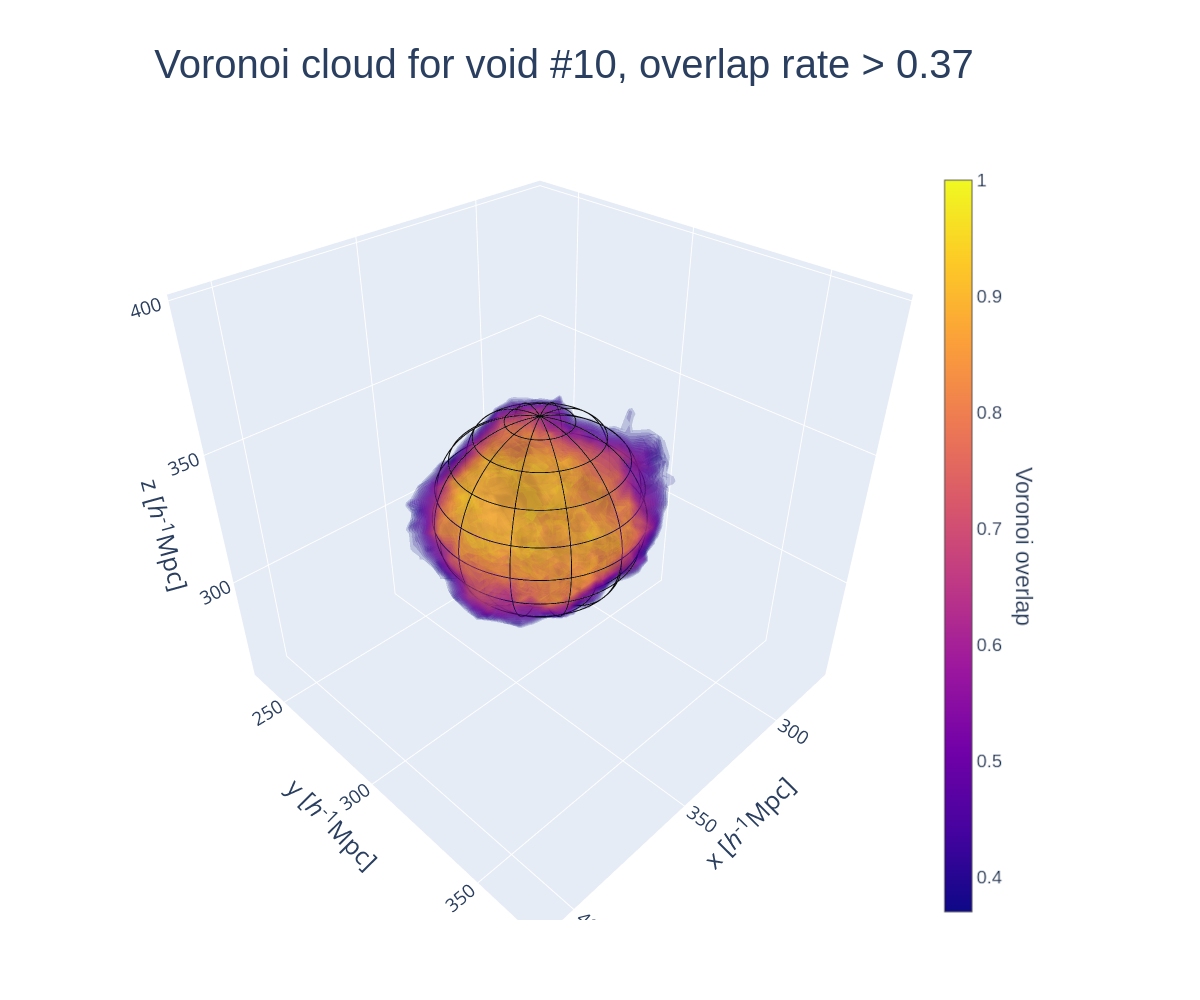}
    \end{minipage}
    \caption{\\
    Left panel: full Voronoi cloud of an example void around its center. The colorscale represents the Voronoi overlap rate, ranging from $0$ to $1$, i.e. how likely a point is to belong to the void throughout different realizations. This value is loosely related to the actual underlying probability distribution, with the yellow regions corresponding to the innermost underdense environment. The wireframe sphere in the middle represents the effective volume of the void, and the Voronoi cloud extends significantly beyond this region. \\
    Right panel: Voronoi cloud truncated to make values of Voronoi overlap rate greater than $0.37$, as interpolated from the linear fit of Figure~\ref{fig:linear_rad} in Appendix~\ref{app:cloud_truncation}. The outskirts are removed, and the remaining cloud conserves the mean volume inferred from the Bayesian analysis, while still retaining information on the shape of the void.
              }

    \label{fig:VoronoiCloud}
\end{figure*}

\subsection{Void clustering algorithm}
\label{subsec:clustering}

As mentioned in the beginning of section~\ref{sec:clustering}, small and large voids detected from independent halo fields might happen to co-occur nearby in space, even without being different realizations of the same true void. 
A straightforward solution to prevent this problem is to sort voids into bins based on their radii, effectively only grouping together comparable ones.

In order to detect clusters among these similarly sized voids, we adopt the \texttt{AgglomerativeClustering} algorithm from the \texttt{scikit-learn} package\footnote{\url{https://scikit-learn.org/dev/modules/generated/sklearn.cluster.AgglomerativeClustering.html}}.
This method belongs to the family of hierarchical clustering algorithms, and employs a bottom-up approach. In the first iteration, all points are clusters of a single member, and only those separated by the shortest distance start being merged.
We choose to use the Ward distance, which minimizes the variance of the points making a cluster. The linkage length is then gradually increased, either until all points are considered as part of the same cluster, or a stopping criterion is defined.
This could be a fixed number of clusters or a maximum threshold distance.
We enforce that the centers of voids should not be farther away from each other than their average radius. This will guarantee that voids in a cluster overlap spatially. We choose the center of the radius bin as the threshold length.

\subsection{Spurious void clusters and acceptance threshold}

The clustering algorithm described so far accepts clusters on the basis of their scatter relative to their average size, yet it does not set any constraint on the number of members. 
Void centers that cannot be merged with many neighbors will make up clusters of just a few members.
If not filtered out, they would be accepted as voids, despite carrying little to no statistical significance.
An acceptance threshold is needed to discard these spurious clusters.

As shown in the left panel of Fig.~\ref{fig:centersOverlap}, the outer part of the simulations is not constrained by data, making the void centers in this region vary freely between different realizations. 
As a result, the union of points from all realizations is spatially uncorrelated in the outer, unconstrained region. We can count the number of excess points that appear in the neighborhood of each individual void center by pure chance, assuming a Poisson process to evaluate the associated probability.
The circle represented in Figure~\ref{fig:centersOverlap} marks a radius of $250 \, \text{Mpc}$, roughly corresponding to the farthest galaxy observed in the 2M++ compilation.
However, we observe some correlated structure even outside the edge of this boundary.
In order to add some buffer, we consider a cube with side $600 \, \text{Mpc} = 408.6 \, \Mpch$ centered on the observer: we use this volume as the interior of the simulation box, and all the voids outside of this region are used to estimate the occurrence of spurious clusters.

Given a particular radius bin, we select the outer voids falling in this range and perform the clustering analysis described in subsection~\ref{subsec:clustering}.
We can then count the occurrence of clusters of $n \ge 1$ points to estimate the $P(n -1)$ Poisson distribution, and compute the value of $n_\text{th}$ corresponding to $P(n_\text{th}-1)<p_\text{th}$, with $p_\text{th}$ an arbitrary threshold.
In order to enforce very high statistical significance, we choose a threshold of $p_\text{th} = 6 \times 10^{-7}$, i.e. the conventional value of $5\sigma$ used in particle physics to claim a detection.
We compute this threshold for different radius bins and reject all the clusters that do not pass this acceptance criterion.
This is a conservative choice that will certainly exclude voids that are reasonably good candidates.
However, the goal of this work is to describe real structures of the Universe, producing a catalog of high-significance voids, albeit not exhaustive.

\subsection{Continuous binning strategy}
\label{subsec:binning_strat}

In addition to the detection of regions in our Local Universe that represent true voids, we are interested in probing the underlying probability distribution of their properties.
Unfortunately, the width of the bins used in the clustering algorithm is an arbitrary choice that affects the voids we actually compare, impacting the estimated variance of the radius probability distribution we would infer from them.
Moreover, for a given choice of bin width, the position of the edges may split a significant cluster, where members of the same true void get grouped into two different clusters that do not pass the significance test individually.

In order to overcome this problem, we fix the width of our radius bins to a value $\Delta R$, producing a window that we shift continuously through the range of radii in our voids sample. 
In such a way, there is always a bin choice that prevents the splitting of significant clusters. 
With this strategy, voids with radii falling into the high density region of the underlying probability distribution belong to clusters that pass the acceptance criterion more often than others. 
Conversely, voids sampled from the tails of the distribution are grouped into smaller clusters that will be labeled as spurious ones.
Counting how often each point is accepted assesses the importance of a single realization void in the evaluation of the statistical properties of the "true" void.
This procedure produces a list of independent voids from different halo field realizations, with their respective weights: we infer the underlying probability distribution with a weighted kernel density estimation (KDE).
We use the \texttt{gaussian\_kde} module from the \texttt{scipy} library\footnote{\url{https://docs.scipy.org/doc/scipy/reference/generated/scipy.stats.gaussian_kde.html}}. 
A toy example of this procedure can be found in Appendix~\ref{app:moving_bin} and Figure~\ref{fig:moving_bin}.

Unfortunately, this procedure is still sensitive to the choice of $\Delta R$.
Too small of a value prevents us from probing the full width of the radius distribution of larger voids, while choosing a too large value might mix together distinct voids that sit nearby, as in the common configuration of smaller satellite voids surrounding a large one. 
As a lower bound, we identify the size of the smallest void in the individual catalogs, $R_\text{min} \sim 5\,\Mpch$. 
Conversely, $\Delta R$ needs to be large enough to probe the width of the underlying distribution of the largest voids. Using the maximum value of the catalog as reference ($R_\text{max} \sim 50\,\Mpch$), we reasonably do not expect the standard deviation to reach half this value.
We choose $25\,\Mpch$ as an upper bound for $\Delta R$, and sample in the middle of the range of $[5,25]\,\Mpch$, repeating the analysis with $\Delta R = 10, 12.5, 15, 17.5, 20\,\Mpch$.

We consider the center value of $\Delta R = 15\,\Mpch$ as reference, and compare the catalog with the ones obtained with the perturbed choices of $\Delta R^{--} = 10\,\Mpch$, $\Delta R^- = 12.5\,\Mpch$, $\Delta R^+ = 17.5\,\Mpch$ and $\Delta R^{++} = 20\,\Mpch$.
As expected, the correspondence is not perfect, with voids disappearing when moving away from the reference catalog obtained with $\Delta R = 15\,\Mpch$.
We then select the intersection of the five catalogs, consisting only of voids that are stable with respect to the choice of $\Delta R$.
This results in a final catalog of $N = 100$ high-significance voids.


\begin{figure}
   \centering
   \includegraphics[width=\hsize]{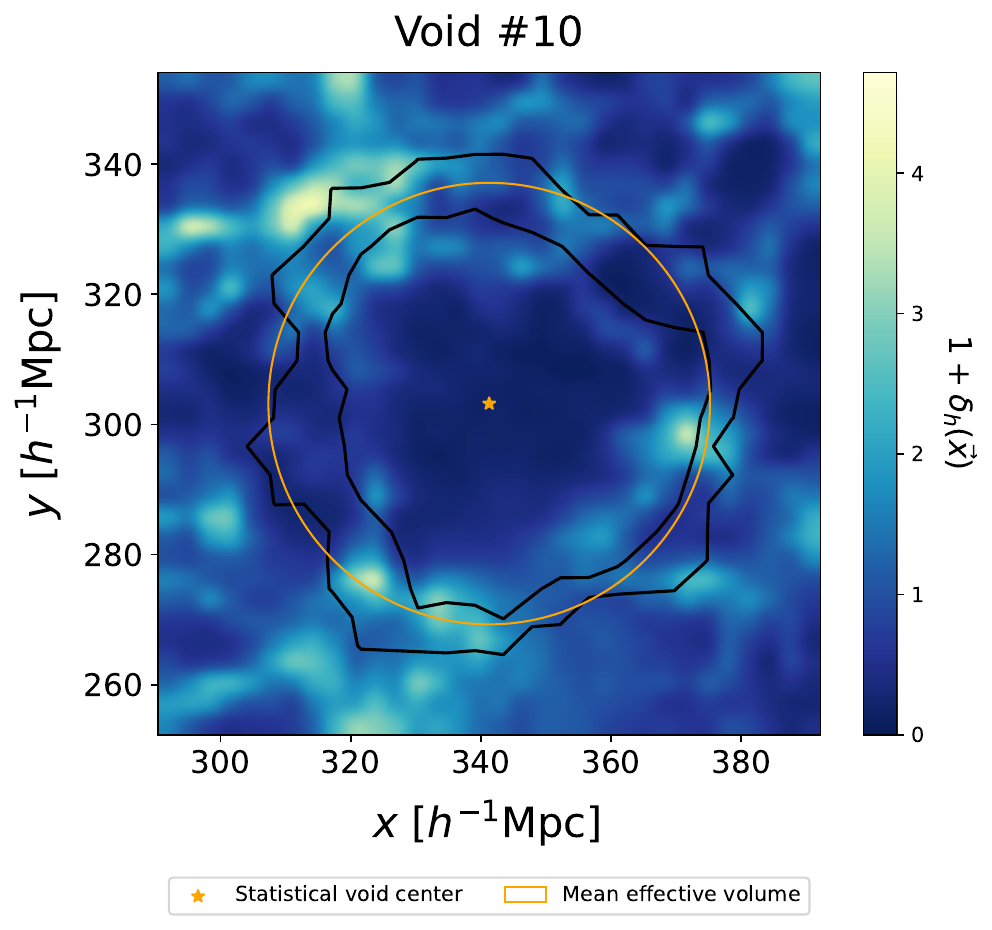}
      \caption{Representation of void \#10 - our best candidate to be the well-known Local void - on a slice of the average halo field with thickness equal to the effective radius $R_\text{void}$. The yellow cross represents the mean posterior of the center, while the circle is as large as the mean effective volume. The black contours represent the truncated Voronoi cloud - i.e. all the points with $r_\text{Voro}>0.37$ - marginalized over the line of sight, obtained by summing the values over the $z$ direction. We normalize this probability distribution to a range between 0 and 1, the contour lines are the 0 and 0.5 levels. The profiles follow perfectly the structure of the underlying field, confirming the quality of the reconstruction. However, the projection does not fully capture the three dimensional information, and heavily depends on the thickness of the density field slab. A representation of different slices on the z-axis can be found at this link: \url{https://voids.cosmictwin.org} \\
      }
         \label{fig:voidContous}
   \end{figure}


\section{Catalog analysis}
\label{sec:analysis}

We produce a catalog of 100 high-significance voids within a distance of $\approx 200 \, \Mpch$, with radii in the $\sim [15 - 45] \, \Mpch$ range.
As a result of the weighting and KDE strategy detailed in section~\ref{subsec:binning_strat}, our catalog includes the probability distributions of position and radii.
We can summarize their properties by computing the mean and standard deviation of these quantities.
The complete results are presented in Appendix~\ref{app:catalog}, while some example voids are shown in Table~\ref{tab:voids_catalog_short}. 
In our cartesian system the $xy$ plane corresponds to the equatorial plane, while the $\hat{z}$ axis point to the equatorial North Pole. The observer is located at $[340.5, 340.5, 340.5]$ \Mpch.
We present the void centers in terms of equatorial sky coordinates ($\alpha$, $\delta$) and distance from us $d_c$.
The mean size $\bar{r}$ of the void and its error is presented in column 6. We use this information to compute the redshift of the closest and farthest edges along the line of sight.
In this section we proceed to analyze the quality of this catalog.

\subsection{Shape characterization}
\label{subsec:shape}

In the discussion presented in Section~\ref{sec:clustering}, we simplified the shape of voids by treating them as spheres.
However, \texttt{VIDE} is a watershed based void finder that performs a Voronoi tessellation on the tracers of the density field and merges them though a watershed transform.
As a result, the output voids are very irregularly shaped. We can use this information to retrieve the actual shape of the statistical voids.

In order to study the morphology of each individual void, we select a box anchored in its center, with side equal to $4 R_\text{void}$, and grid it with resolution $N = 32^3$.
By definition, every point in space is contained in the Voronoi cell corresponding to its closest tracer.
Hence, we flag as one every voxel whose closest tracer is part of a void, repeating the procedure for all the realizations that contribute to that specific void.
We then average them using the appropriate weights introduced in section~\ref{subsec:binning_strat}. 
The result is a value between $0$ and $1$, that we define as the "Voronoi overlap rate" $r_\text{Voro}$, describing how often a point belongs to the Voronoi cells of a realization of a void.
The function defined on the whole space is closely related to the probability that the space around $\vec{x}_c$ is part of the void, and hereafter we will refer to it as a Voronoi cloud.

An example void is presented in Fig.~\ref{fig:VoronoiCloud}. The outskirts of its Voronoi cloud are unlikely to truly belong to the void, as they were identified as part of a void only in fewer halo field realizations.
For all voids, we reduce the size of the clouds by truncating the points of the grid with Voronoi overlap rate of $r_\text{Voro} > r_\text{min}$, testing different thresholds $r_\text{min}$.
We evaluate the volumes of these sub-clouds, we convert them into the corresponding radii, and we compare them to the mean posterior of the radii distributions obtained from the KDEs.
We observe that the dependence between the two quantities is linear, as detailed in Appendix~\ref{app:cloud_truncation} and shown in Fig~\ref{fig:linear_rad}.
In particular, choosing $r_\text{min} = 0.37$ we find that the ratio between the two definitions of void radius equals $1.0004 \sim 1$ within sub-percent accuracy.
We adopt this threshold in order to determine whether a point belongs to a void, effectively characterizing the actual shape of the void, without overestimating its volume. 

We can finally analyze the spatial interplay and connections between these individual voids by painting their Voronoi clouds onto the whole volume.
We define a grid of $64^3$ voxels, with coarser resolution of $\sim 7.4\,\Mpch$.
We then assign each point of the single Voronoi clouds to the closest voxel, averaging the multiple contributions in the cases where multiple voxels of an individual void cloud fall into a single voxel of the global grid.
We then choose $r_\text{Voro} = 0.37$ as the threshold that conserves the mean volume of all voids within sub-percent error, yielding a ratio of $0.995$. The global Voronoi cloud for all voids is shown in Fig.~\ref{fig:allVoids}.

\subsection{Average properties}
\label{subsec:average_properties}

In order to develop the optimal void clustering strategy for our problem, we carried out a blinded analysis without adjusting our choices to the underlying halo fields.
We now compare the statistical voids of our catalog to the mean halo distribution to check their consistency. 
We assess the quality of our detection through visual inspection.
For each void, we select the truncated Voronoi cloud with $r_\text{Voro}>0.37$, we sum the overlap rate on the z direction, and we normalize the result to have a value between $0$ and $1$.
This is qualitatively equivalent to marginalizing the probability distribution over the $z$ direction.
We then plot a slice of the halo field averaged over all realizations, with thickness equal to $R_\text{void}$, and we overlap the contours corresponding to $0$ and $0.5$ marginal Voronoi overlap rate.
In Fig~\ref{fig:voidContous} we show an example void.
We find a good qualitative match between the void profile and its density environment, as the void is not only centered in an under-dense region, but the contours follow the shape of the underlying structures. 

\begin{figure}

    \centering
    \includegraphics[width=\linewidth, trim={0.2cm 0cm 3.7cm 0cm},clip]{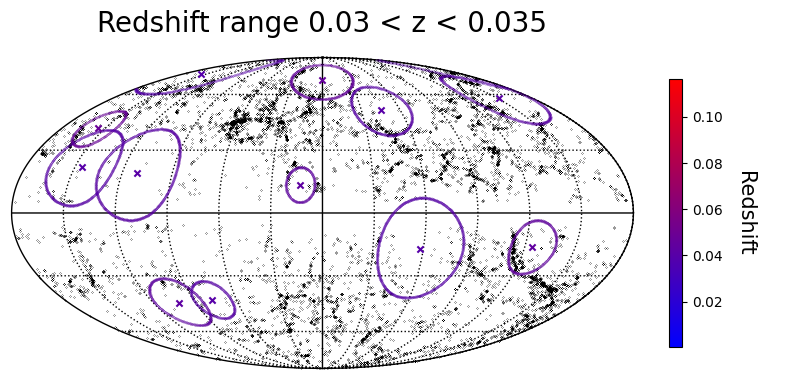}
    
    \caption{\\
    Sky representation in galactic coordinates, with the zero longitude point in the center, with redshift $0.03<z<0.035$. The crosses represent the void centers, while the circles are the apparent size of their effective radii. The black dots represent the 2M++ galaxies used to infer the posterior realizations of the Local Universe. As a visual confirmation, voids actually occupy the regions with little to no galaxies. The predictive power of the \texttt{Manticore-Local} suite combined with our void finding procedure is shown by the fact that we are able to infer voids behind the galactic plane, in regions where we do not have direct observations of galaxies. The remaining redshift slices are shown in Figure~\ref{fig:sky_projection_all} in Appendix~\ref{app:catalog}.
              }

    \label{fig:sky_projection_single}

\end{figure}

Finally, we can visualize the full catalog using the spherical approximation.
In Fig~\ref{fig:sky_projection_single} we represent the sky at redshift $0.03 < z < 0.035$, representing the galaxies of the 2M++ compilation and the void centers on top of it.
The circles represent the angular size of the effective radius projected on the sky at the center's redshift. The remaining redshift bins are shown in Figure~\ref{fig:sky_projection_all} in Appendix~\ref{app:catalog}.

\section{Comparison with literature voids}
\label{sec:comparison}

In this section, we discuss how our catalog compares to the known voids that have been historically identified by astronomers with observational methods.
Such task is non-trivial, as voids not only subtend great angles in the sky, but they are also interconnected, making their centers and boundaries not well defined.
They are often named after the constellation that lies in front of them for practical purposes, but there is no indisputable catalog of voids the community agrees upon.
We proceed in a qualitative comparison with previous works.
The voids from our catalog mentioned in the following sections are all summarized in Table~\ref{tab:voids_catalog_short}.

\subsection{Dynamical void detection}
\label{subsec:cosmicflows}

\citet{Tully2019} have characterized the shape and interplay of voids, by analyzing the velocities of the surrounding galaxies to reconstruct the underlying density field, using the \textit{Cosmicflows-3} dataset \citep{Tully2016}.
We compare with the 25 local minima they identify, classified as part of the Local void, the Hercules void, the Sculptor void, and the Eridanus void. 
We first convert their coordinates into our simulation volume, and evaluate the Voronoi cloud in these points.
We consider the Voronoi overlap rate of the eight closest voxels of the grid representation, and average them using the inverse of the distance as weights.
If the estimated value is higher than the acceptance threshold of $0.37$, we flag the candidate point as part of a void.
We find that $17$ out of $25$ pass the acceptance threshold, corresponding to a $68\%$ confirmation rate. 

As a first cross-check, we compare the remaining eight points to the mean halo field, and we do not find significant overdensities overlapping to those regions. 
Out of the rejected points, seven present a low value of the Voronoi overlap rate ($\lesssim~0.2$), and are located at a significant distance from our global Voronoi cloud. 
These points are still likely to be their own independent underdensities, but their significance is lower than our designated $5\sigma$ threshold.
The remaining point - named "Lacerta-2.4" in the paper - sits close to the boundaries of the global Voronoi cloud, at a distance of a few \Mpch, despite not strictly belonging to any void of our catalog.
These differences can be due to the $-\nabla \cdot\vec{v}$ estimation of the density, as the two fields de-correlate at small scales.
Taking these differences into account, the agreement is overall good.

As far as it concerns the Local void, we find good agreement for the point marked as "Aquila-0.8" point, which is part of void \#10 of our catalog.
The "Andromeda-2.3" point is fully inside void \#75 of our catalog, while the aforementioned "Lacerta-2.4" is located close to its edge.
This spatial distribution matches the full visualization of the void in \citet{Tully2019}.
This comparison highlights how what is commonly referred to as the Local void is the connection of separate voids.

\subsection{The Local Void}
\label{subsec:local}

The Local Void is hard to characterize, as it partially lies behind the galactic plane and subtends a big portion of the sky due to its vicinity.
First discovered by \citet{TullyFisher}, different works have defined different positions for its center.

\citet{Karachentseva1999, Karachentsev2002} locate its center at $\alpha = 18^\text{h}38^\text{m}$, $\delta = 18^\circ$, and redshift range $< 1 \ 500 \ \text{ km s}^{-1}$;
while \citet{Nakanishi1997} estimate the center of the Local void at $l \sim 60^\circ$, $b \sim -15^\circ$ (i.e. $\alpha \sim 20^\text{h}40^\text{m}$, $\delta \sim 16^\circ$), its distance at $cz \sim 2 \ 500 \ \text{km s}^{-1}$, and the size of the void along the line of sight as $cz \sim 2 \ 500 \ \text{km s}^{-1}$.
As the Local void's boundaries are not unequivocally defined, different authors happen to use the same name to refer to different regions.
\citet{Karachentseva1999} point out the vicinity of these two separate voids, that can thus be confused as the same Local void, and distinguish them by calling them, respectively, the "Hercules-Aquila Void” and the “Pegasus-Delphinus Void”. 
However, we do not find this separation in our catalog, as both of them overlap with void \#10, whose center is located at $\alpha \sim 18^\text{h}05^\text{m}$, $\delta \sim -16^\circ$, and redshift $3 \ 900 \text{ km s}^{-1}$, as shown in Table~\ref{tab:voids_catalog_short}.
With its center at a distance of $\sim 39\,\Mpch$ and radius of $\sim 34\,\Mpch$, this void is confirmed to be our best candidate of the Local Void.

Additionally, \citet{Einasto1994} define the Northern Local Void (NLV) as centered in $\alpha = 256.1^\circ$, $\delta = -4.8^\circ$, at a distance of $61\,\Mpch$ away from us, and with a radius of $52\,\Mpch$.
Void \#10 of our catalog overlaps perfectly within its range, but it only extends to a distance of $ \sim 73\,\Mpch$. Towards the farthest edge of this region, we find void \#83, with a smaller radius of $ \sim 14\,\Mpch$ and connected to void \#10 through a bridge.
The volume claimed by the authors is much bigger than other accounts of the Local void, and it extends farther away from us. 
However they looked for empty regions in the distribution of clusters of galaxies, which are a much sparser tracer of the desnity field than the galaxies themselves: as finer structures like filaments are not resolved, smaller adjacent voids appear to be a bigger one.
As opposed to the Northern Local Void, \citet{Einasto1994} also mention a Southern Local Void (SLV), centered in $\alpha = 121.7^\circ$, $\delta = -1.5^\circ$, at a distance of $96\,\Mpch$ away from us, and with a radius of $ 56 \ \,\Mpch$. However, we do not find any significant underdensities in that direction.

\subsection{Other voids}
\label{subsec:other_voids}

Historically, the first void detection is attributed to \citet{Gregory1978}, who noticed a region devoided of galaxies behind the Coma/A1367 supercluster. Void \#45 in our catalog matches the sky coordinates, and the redshift range of $5 \ 000-6 \ 200\  \text{km \ s}^{-1}$, as shown in Table~\ref{tab:voids_catalog_short}.
Early records of voids in the Local Universe can be found in \citet{Kirshner1981}. Void \#88 of our catalog is centered in the Boötes constellation~\footnote{
Unless the reference explicitly specifies a coordinate range, in order to determine in which constellation our void centers are located we use the IAU defined boundaries through the \texttt{astropy.coordinates} function \texttt{get\_constellation()}: \url{https://docs.astropy.org/en/stable/api/astropy.coordinates.get_constellation.html}
}
and extends on a redshift range of $10 \ 000-18 \ 100\  \text{km \ s}^{-1}$, overlapping with the void claimed by the authors.
Similarly, the Pisces void marginally matches void \#15, while void \#75 extends in the Perseus/Pisces region defined in the reference.
It is worth noting that the latter is very nearby, at a distance of $ \sim 33\,\Mpch$, and in fact it can be considered part of the Local void of \citet{Tully2019}, as described in section~\ref{subsec:cosmicflows}.

The work by \citet{Kirshner1981} also mentions a void in the Hercules constellation, in the redshift range of $5 \ 500-8 \ 500\  \text{km \ s}^{-1}$, while \citet{Freudling1991} studied a very nearby region, with ranges $14^\text{h} < \alpha < 17^\text{h}$, $10^\circ < \delta < 60^\circ$ and redshift $4 \ 900-8 \ 900\  \text{km \ s}^{-1}$;
conversely, \citet{Krumm1984} define a very narrow Hercules region ($15^\text{h}45^\text{m} < \alpha < 16^\text{h}15^\text{m}$, $14^\circ < \delta < 22^\circ$) at approximately the same redshift range, as well as claiming a Perseus void at $0^\text{h} < \alpha < 2^\text{h}$, $0^\circ < \delta < 20^\circ$ and redshift $6 \ 500-9 \ 500\  \text{km \ s}^{-1}$.
We do not find high-significance voids in any of those regions.

Other named voids can be found in \citet{Willmer1995}, who report the Hydra void in the region centered in $\alpha = 11^\text{h}$, $\delta = -30^\circ$, and redshift range $4 \ 500-6 \ 000\  \text{km \ s}^{-1}$.
This volume corresponds to our void \#57, which however extends up to redshift $\sim 9 \ 400 \ \text{km s}^{-1}$. The same authors mention the Leo void at $\alpha = 11^\text{h}30^\text{m}$ and redshift $4 \ 000 \ \text{km s}^{-1}$ in the declination range of $-10^\circ < \delta < 10^\circ$, which overlaps perfectly with void \#28 of our catalog.
Finally, \citet{Pustilnik2006} studied the Pegasus void as defined by \citet{Fairall1998}, with coordinates of the center in $\alpha = 22^\text{h}$, $\delta = 15^\circ$, and $cz = 5 \ 500 \ \text{km s}^{-1}$. However, we do not find a match with any of our voids in the range of $cz = 3 \ 000 \ \text{km s}^{-1}$ mentioned by the authors.

\subsection{Final remarks on the comparison}

In conclusion, the centers and boundaries of voids in surveys are not always precisely defined, hence our comparison is mostly qualitative.
In the literature voids are often identified as regions with little to no galaxies, which tend to be smaller in size compared to voids in halos or dark matter, as the latter still contain some galaxies.
Overall, we find a good match with the known voids discovered observationally by astronomers. However, our catalog is not exhaustive, as our acceptance threshold is very high due to the $5\sigma$ criterion: observed voids that do not match our work are not disproved, as they are likely to be real but slightly less significant. Conversely, the strictness of our criterion makes us confident that the voids we claim are real, and can be used for any applications that needs a characterization of the density environment of astronomical objects. Finally, for practical reasons our comparison is with respect to a limited selection of works on observational voids, and as such we are aware of potentially missing relevant sources.

\section{Discussion and conclusions}
\label{sec:conclusion}

In this work, we presented a catalog of high-significance cosmic voids extracted from constrained simulations of the Local Universe. For our analysis we used the \texttt{Manitcore-Local} suite from \citet{manticore}, a set of 50 N-body simulations initialized with the primordial density field inferred with the \texttt{BORG} algorithm \citep{borg,Jasche2019}.
Produced in a Bayesian framework, they represent independent statistical realizations of the posterior distribution of the cosmic web.
To characterize the predicted structures, we ran the \texttt{VIDE} void finder \citep{vide} on the halos extracted from each individual realization, obtaining a set of 50 independent void catalogs.
We overlapped the centers of these voids on the three-dimensional space, finding notable clusters of points in particular locations, as shown in Figure~\ref{fig:centersOverlap}.
These clusters correspond to the regions that are most likely to be real voids in the dark matter distribution.

We developed the optimal clustering strategy to identify the void clusters in an unbiased way, avoiding arbitrary choices of free parameters.
We evaluated the statistical significance of each detection, modeling the probability of void clusters emerging in the union of uncorrelated realizations of the large-scale structure through a Poisson process.
We rejected the spurious ones by imposing a $5\sigma$ detection threshold, i.e. accepting only the clusters that could have occurred by random chance with  $\sim 6 \times 10^{-7}$ probability.
The result of this work is a catalog of $100$ voids, consisting of posterior distributions of their centers positions and radii.
Their main properties can be found in Appendix~\ref{app:catalog} for the complete catalog, with some notable voids summarized in Table~\ref{tab:voids_catalog_short}. 
This statistical result only considers the effective volume of voids, which is equivalent to approximating them as spheres. 
However, the \texttt{VIDE} algorithm finds voids through a Voronoi tesselation of the tracers of the density field, identifying cells to be merged into zones and voids using the watershed transform \citep{Platen2007}, yielding very complex morphologies.
We used this information to reconstruct the full shape of voids, as presented in Figure~\ref{fig:VoronoiCloud} for an example void, and in Figure~\ref{fig:allVoids} for the whole catalog. Further visualizations can be found in the dedicated website\footref{website_url}, while the catalog is publicly available for download\footref{catalog_url}.

\bigskip

To check the quality of the voids in our catalog, we compared them directly to the mean posterior of the underlying halo field.
We find good overall consistency between the profile of the voids and the density environment, as shown in Figure~\ref{fig:voidContous} for an example void.
Hence, the reliability of the catalog is strongly dependent on the quality of the \texttt{Manticore-Local} simulations and its posterior predictions.
The reconstruction of the initial conditions of the Universe through \texttt{BORG} is now a well-established tool for field-level cosmology.
Ever since its first publication by \citet{borg}, the algorithm has seen notable improvements, passed several consistency tests, and has undergone continuous refinements to remain state-of-the-art technology.
We refer to \citet{Jasche2015, Lavaux2015, Jasche2019, Lavaux2019} for further details.
As a second check, the \texttt{Manticore-Local} simulations were carefully tested to be consistent with the cosmological model, to be able to predict known galaxy clusters, and to reconstruct the velocity field, as detailed in \citet{manticore}.

Possible sources of bias in the procedure can occur if the chains are not fully converged and the samples are correlated.
These issues would particularly have an effect on the outer regions of the box, which are not constrained by data, where the Gaussian prior on the initial conditions is the predominant element of the inference.
\citet{manticore} addressed this possibility by running five independent MCMC chains passing the Gelman-Rubin convergence test.
In addition, the 50 samples of the final \texttt{Manticore-Local} set are taken at sufficient separation along the chains to be uncorrelated. 
As a further confirmation, the right panel of Figure~\ref{fig:centersOverlap} shows that the exterior of the box does not present significant structures - with density contrast close to zero - visually illustrating that the MCMC chains converged.
For our purposes, the samples can be safely treated as independent realizations of the posterior distribution of the large-scale structure.

One possible limitation of this work is the use of a single void finding algorithm, instead of verifying the consistency of the results with respect to different choices.
However, as discussed in section~\ref{subsec:vide}, \texttt{VIDE} is not only a well-established, carefully tested, and widely used tool in the void literature, but it also presents desirable features, such as shape agnosticism and high computational performance on simulations.
Our void clustering strategy only considers first order properties of voids, i.e. center positions and sizes, which are overall consistent among void finders, at least for the largest voids \citep{Colberg2008}.
It is reasonable to expect comparable results for different void definitions: possible outlier voids emerging due to the specific characteristics of a particular algorithm would end up being rejected by the $5\sigma$ detection threshold.
The statistical nature of the procedure - inferring posterior distributions from independent realizations - should attenuate the differences between void finders, making the catalog robust to these variations; however, these claims need to be tested in future works.

Finally, we compared our catalogs with different reports of observational detections of voids in the literature.
We considered underdensities inferred from velocity measurements in \citet{Tully2019}, as well as voids identified as regions with few to no galaxies.
Taking into account the intrinsic challenges of the comparison - i.e. the fundamental differences of the procedures, and the not well-defined centers and boundaries - we find overall good agreement with the references we considered.
It is worth pointing out that our catalog only contains voids with very high statistical significance, and as such it is not exhaustive: we are aware that we are certainly missing real voids due to the strictness of our selection, and as such we are not disproving other accounts of known voids that do not appear in our catalog.

\bigskip

In conclusion, this work introduces a viable benchmark to identify cosmic voids in the Local Universe, and to assess their statistical significance, producing a catalog of voids that are real structures in the dark matter distribution, as opposed to artifacts of the galaxy surveys.
Our method allows us to infer the posterior distributions of void properties in a Bayesian framework, obtaining an estimation of their statistical uncertainties.
Additionally, our voids have well-defined centers, shapes, and boundaries, making the catalog practical to use for all applications that need characterization of the density environment.
This includes galaxy properties and their evolution, galaxy biasing, supernovae and their use for Hubble constant measurements, and high-energy astrophysics, such as black holes and AGN in voids.
We make the catalog publicly available\footref{catalog_url}
and encourage researchers from different fields of astronomy to download it and use it for their work.

Ongoing efforts within the Aquila consortium will produce constrained simulations on larger volumes, named \texttt{Manticore-Deep}, leveraging the information from a larger galaxy survey such as the Sloan Digital Sky Survey \citep{SDSS}.
We expect to extract a larger number of voids from these suites due to volume scaling, characterizing the deeper Universe, and increasing the statistical power of this probe. 
This can be used for cosmology through cross-correlations with the large-scale structure through the weak gravitational lensing of galaxies, or with the Cosmic Microwave Background (CMB).
Possible tests for this latter case include the late-time integrated Sachs-Wolfe (ISW) effect, the thermal and kinematic Sunyaev-Zel'dovich (SZ) effect, or a direct comparison with the density maps reconstructed from cosmic shear.

\section*{Data availability}

The \texttt{Manticore-Local} constrained simulations will be publicly available\footnote{\url{https://cosmictwin.org}}.
We provide supplementary visualizations for this work on the public website\footnote{\label{website_url}\url{https://voids.cosmictwin.org}}, while the derived void catalog can be downloaded\footnote{\label{catalog_url}\url{https://github.com/RosaMalandrino/LocalVoids/}} and used by the community.

\begin{acknowledgements}
      We thank Paul Sutter, Alice Pisani, Florent Leclercq, Clotilde Laigle, and R. Brent Tully for useful comments and discussions.
      This work has been done within the Aquila Consortium\footnote{\url{https://www.aquila-consortium.org/}} and the Simons Collaboration on "Learning the Universe"\footnote{\url{https://learning-the-universe.org/}}
      This work has made use of the Infinity Cluster hosted by Institut d’Astrophysique de Paris; the authors acknowledge St\'ephane Rouberol for his efficient management of this facility.
      This work benefitted from the HPC resources of TGCC (Très Grand Centre de Calcul), Irene-Joliot-Curie supercomputer, under the allocations A0170415682 and SS010415380.
      JJ, GL, and SM acknowledge support from the Simons Foundation through the Simons Collaboration on "Learning the Universe". This work was made possible by the research project grant "Understanding the Dynamic Universe," funded by the Knut and Alice Wallenberg Foundation (Dnr KAW 2018.0067). Additionally, JJ acknowledges financial support from the Swedish Research Council (VR) through the project "Deciphering the Dynamics of Cosmic Structure" (2020-05143). BDW acknowledges support from the DIM ORIGINES 2023 "INFINITY NEXT" grant. The Flatiron Institute is a division of the Simons Foundation.
\end{acknowledgements}

%
%

\bibliographystyle{aa} 
\bibliography{biblio} 

\begin{appendix} 
\let\part=\section\appendix
\onecolumn

\section{Continuous binning toy example}
\label{app:moving_bin}

\begin{figure*}[]
    \centering
    \includegraphics[width=\textwidth]{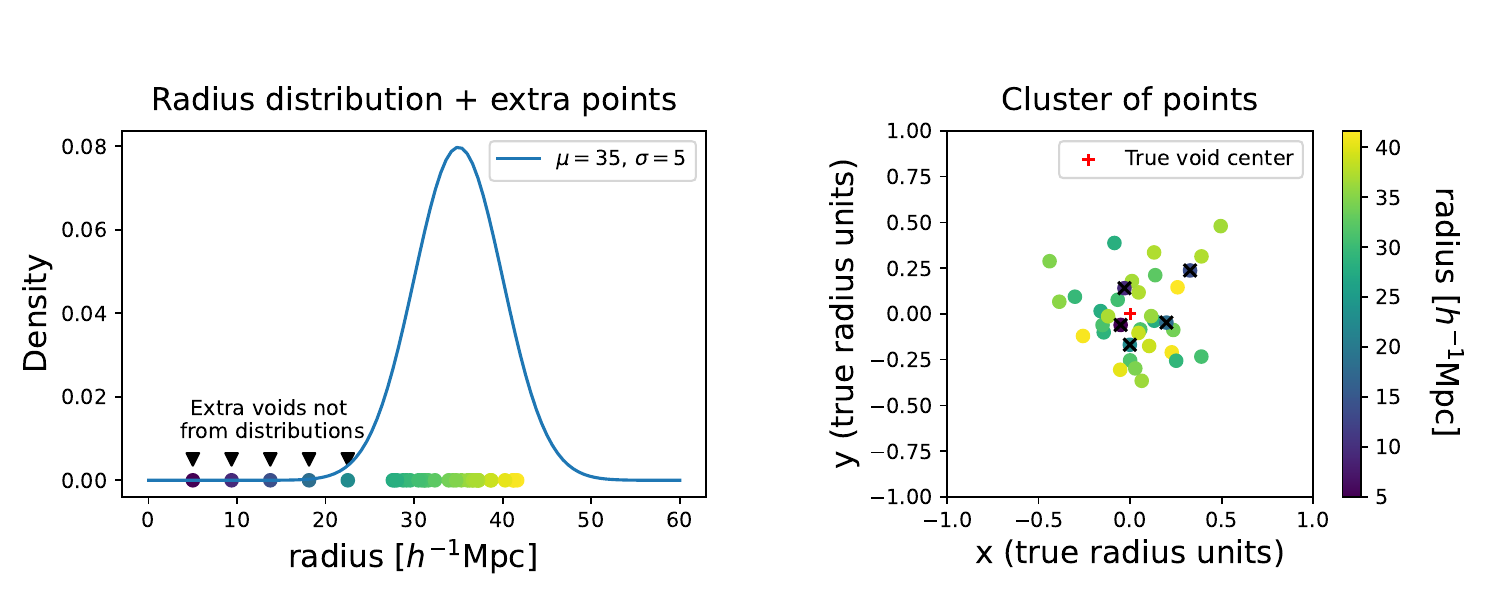}
    \includegraphics[width=\textwidth]{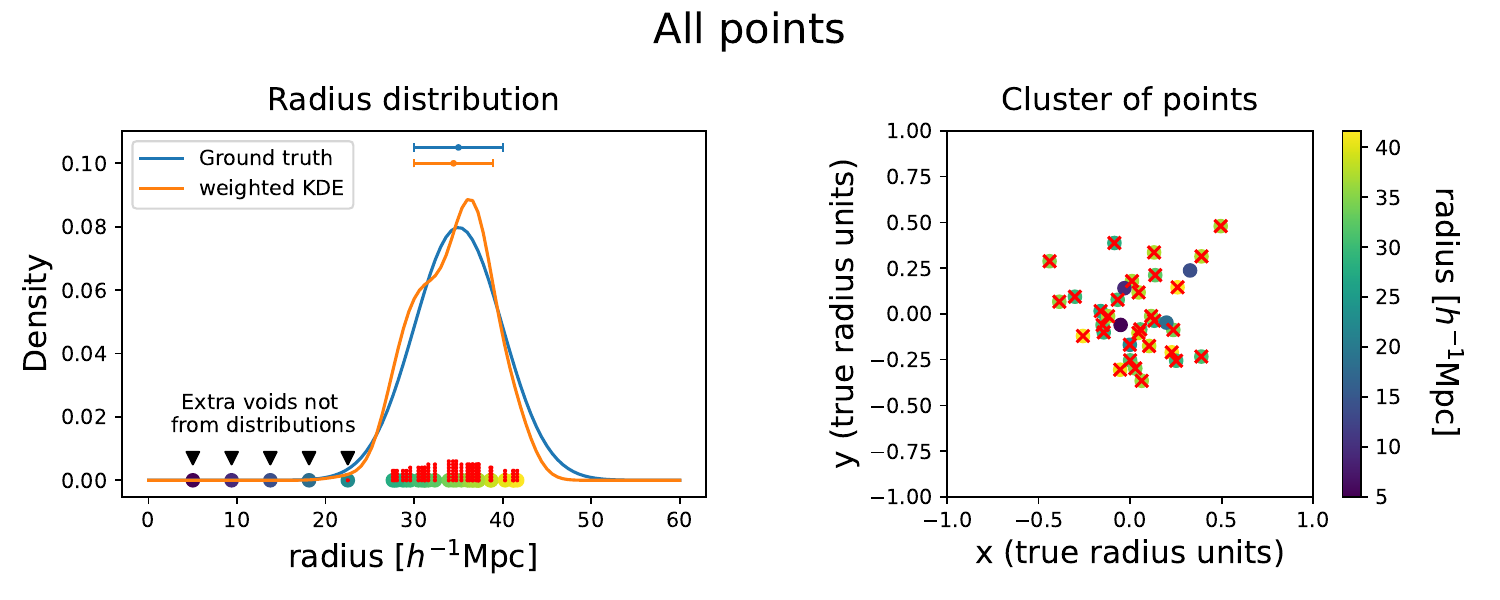}
    
    \caption{Toy model illustrating the continuous binning strategy.
            The top left panels represents samples of a radius distribution, the top right of the position around the true center. Some nuisance smaller voids are added.
            The bottom panels represent the result of the technique, with the probability distribution of the radius estimated through a weighted kernel density estimation (KDE).
            Despite some differences, we find a good match in the mean and standard deviation of the ground truth vs estimate distributions (represented with the error bars on top).
            A full animation of the procedure can be found here: \url{https://voids.cosmictwin.org/VoidClustering}.
    }
    \label{fig:moving_bin}
\end{figure*}

In Figure~\ref{fig:moving_bin} we present a toy model that illustrates the continuous binning strategy to estimate the probability distributions of the properties of voids, as introduced in Section~\ref{subsec:binning_strat}.
The top panels represent samples from the underlying distribution of the properties of a single voids. On the left, we sample radii from a Gaussian distribution around a true size of $35 \ \Mpch$, while on the right we simulate some scatter around the void's true position.
We add some nuisance smaller voids that might occur nearby in some realizations as an effect of statistical fluctuations and to void fragmentation due to the void finder.
The bottom panels represent the result of the technique, with the red dots counting how often the point has been accepted as part of the statistical voids. These occurrences can be used as weights in a kernel density estimation (KDE) to infer the underlying distribution.
The resulting curve presents some difference, but the true values of its summaries, such as mean and standard deviation, are reconstructed, as represented by the error bars on top.
The technique is overall able to avoid spurious voids: it might still count the ones located at the tails of the distribution, but they do not distort the inference significantly.

\section{Voronoi clouds truncation}
\label{app:cloud_truncation}

\begin{figure*}[t]
    \centering
    \includegraphics[width=\textwidth]{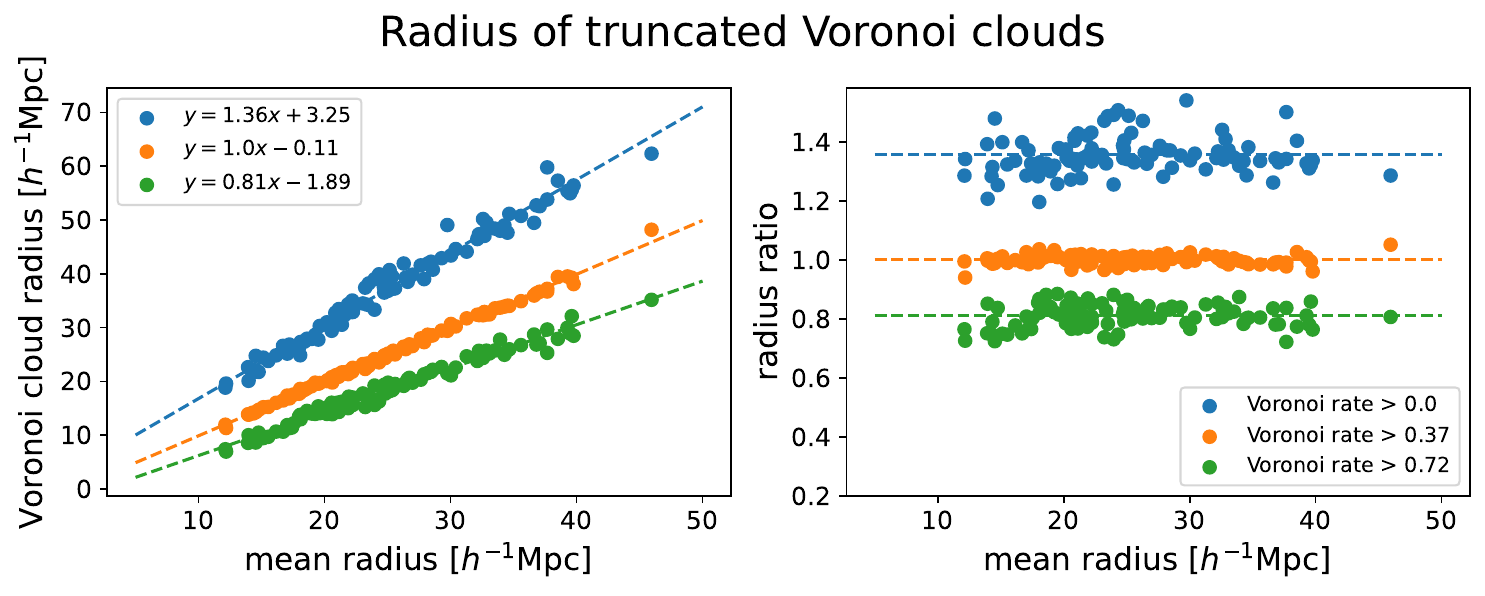}
    
    \caption{Left panel: mean radius inferred from the posterior distribution vs effective radius of the corresponding Voronoi cloud with different truncation thresholds of $r_\text{Voro}>0; \ 0.37; \ 0.72$. Right panel: ratio between the radii estimated in the two different ways, after subtracting the $y$ axis offset, which is in the order of a few \Mpch for all cases. \\
    The relation is linear for all three cases, and provides information on the threshold to impose in order to achieve a certain depth in the void profile.}
    \label{fig:linear_rad}
\end{figure*}

\begin{figure*}[t!]
    \centering

    \includegraphics[width=\textwidth]{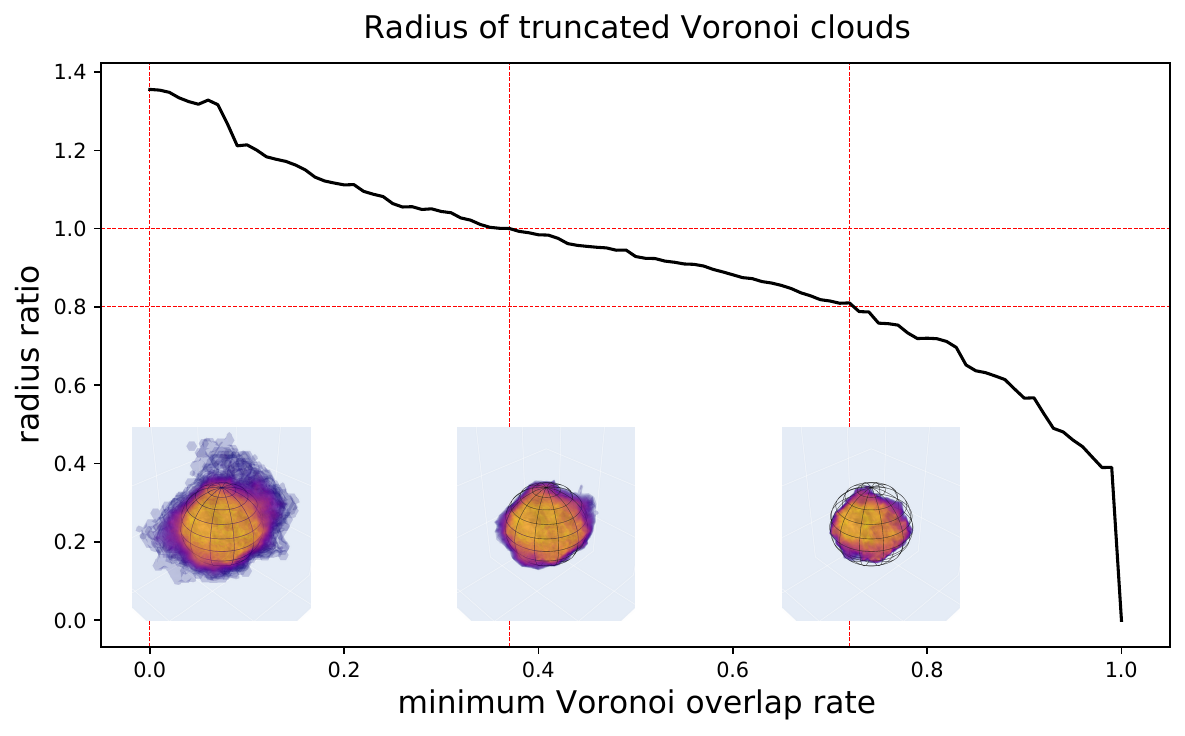}

    \caption{Radius of the truncated clouds with respect to the statistical radius of the voids, as a function of the Voronoi overlap rate threshold. The small panels illustrate the cloud at different truncation levels: when no threshold is applied, the outskirts are larger than the effective volume represented with the wireframe sphere; a minimum overlap rate of $0.37$ preserves the statistical volume, while larger thresholds yield the very interior of the voids. We illustrate the $ \sim 80\%$ selection, corresponding to $r_\text{Voro} > 0.72$, but deeper cuts can be applied if one wishes to probe the most underdense regions.
              }

    \label{fig:Voronoi_cutoff}

\end{figure*}

In this section, we discuss how the Voronoi clouds introduced in Section~\ref{subsec:shape} relate to the actual void shapes, and the necessary truncations.
We compute the volume of each Voronoi cloud by summing the volume of all voxels with $r_\text{Voro}>r_\text{th}$ for different values of $r_\text{th}$ ranging from zero to one. In Figure~\ref{fig:linear_rad} we present the relation between the effective radii of the truncated Voronoi clouds and the mean posterior of the radius distributions, for some example values of $r_\text{th}$.
The relation is linear for all three cases, and provides information on the threshold to impose in order to achieve a certain depth in the void profile.
With truncation at $r_\text{th} = 0.37$ the Voronoi cloud preserves the statistical volume, while a threshold of $0.72$ provides voids of $ \sim 80\%$ the size of the result of the statistical method.
This piece of information can be useful to select the deeper interior of the voids.
In Figure~\ref{fig:Voronoi_cutoff} we present the complete dependency between the truncation level $r_\text{th}$ on the $x$-axis and the size of the corresponding cloud, represented by the average ratio with respect to the mean posterior estimates.

\section{Complete catalog}
\label{app:catalog}

\begin{figure*}
    \begin{minipage}[t]{0.33\linewidth}
        \flushleft
        \vspace{1cm}
        \includegraphics[width=\linewidth, trim={0.2cm 0cm 3.7cm 0cm},clip]{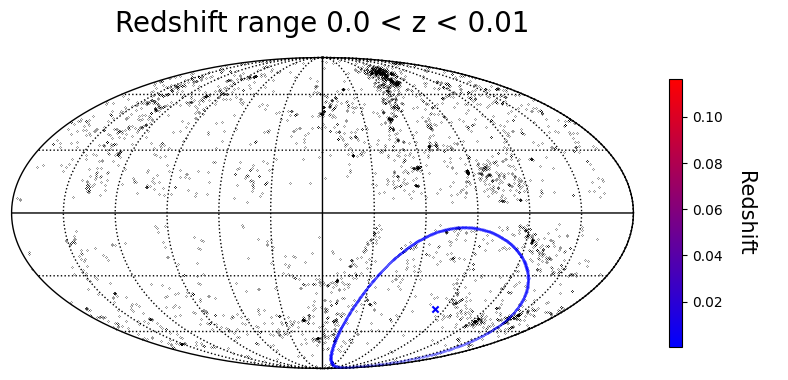}
        \\
        \vspace{0.5cm}
        \includegraphics[width=\linewidth, trim={0.2cm 0cm 3.7cm 0cm},clip]{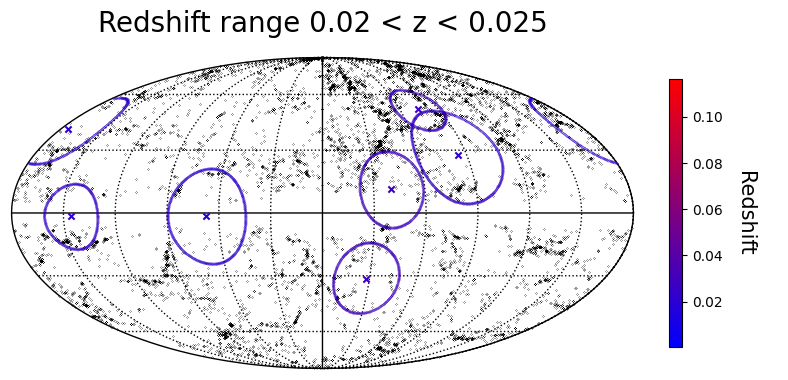}
        \\
        \vspace{0.5cm}
        \includegraphics[width=\linewidth, trim={0.2cm 0cm 3.7cm 0cm},clip]{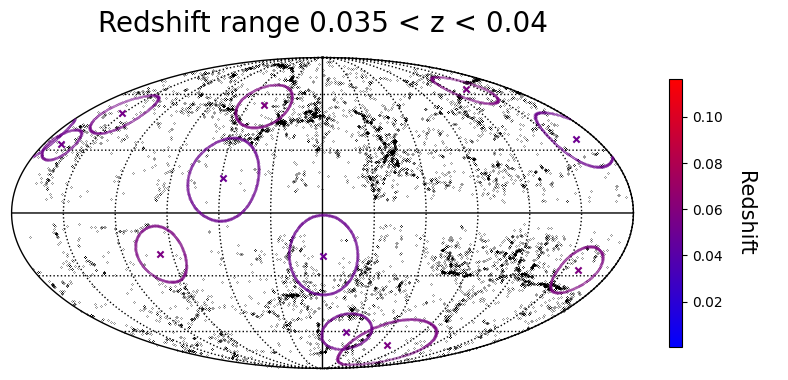}
        \\
        \vspace{0.5cm}
        \includegraphics[width=\linewidth, trim={0.2cm 0cm 3.7cm 0cm},clip]{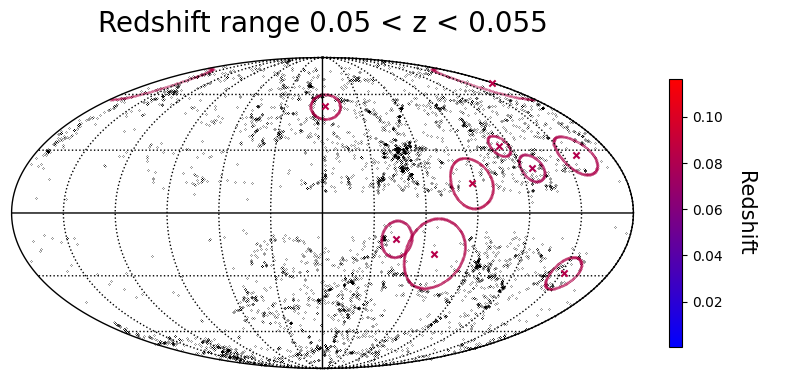}
        \\
        \vspace{0.5cm}
        \includegraphics[width=\linewidth, trim={0.2cm 0cm 3.7cm 0cm},clip]{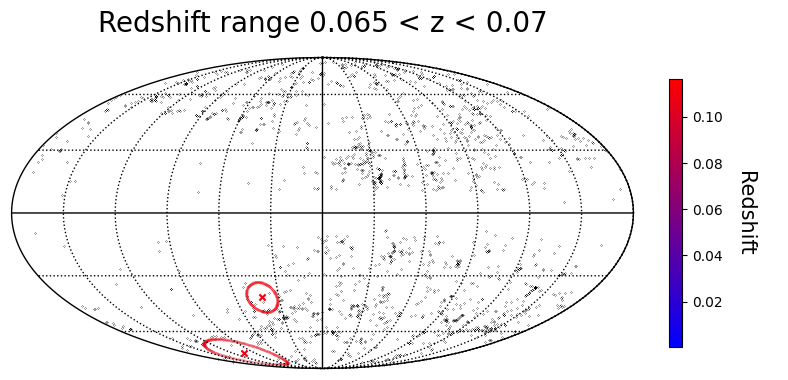}   
        \\
        \vspace{0.5cm}
        
    \end{minipage}
    \begin{minipage}[t]{0.33\linewidth}
        \centering
        \vspace{1cm}
        \includegraphics[width=\linewidth, trim={0.2cm 0cm 3.7cm 0cm},clip]{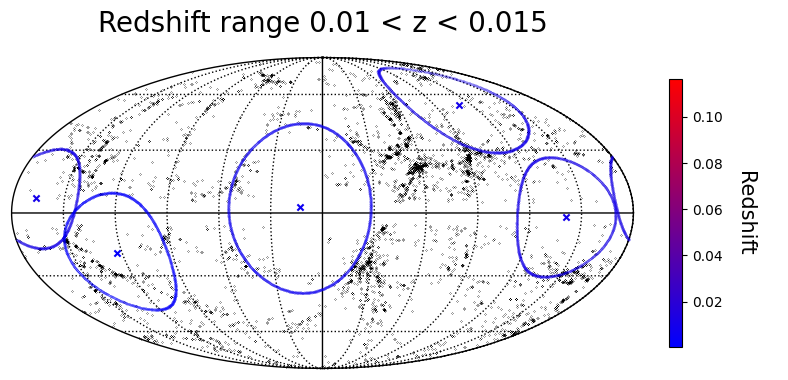}
        \\
        \vspace{0.5cm}
        \includegraphics[width=\linewidth, trim={0.2cm 0cm 3.7cm 0cm},clip]{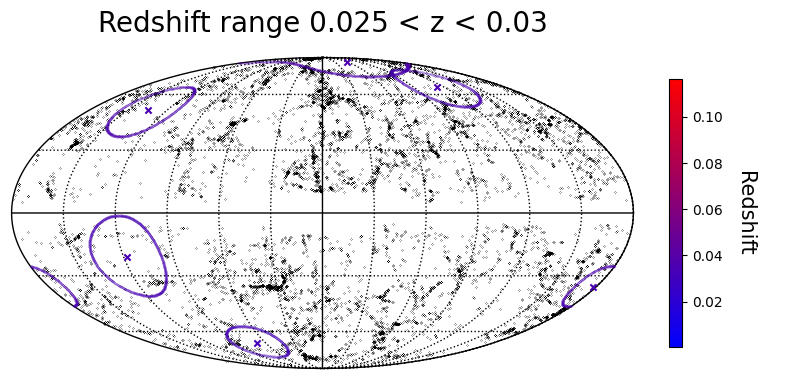}
        \\
        \vspace{0.5cm}
        \includegraphics[width=\linewidth, trim={0.2cm 0cm 3.7cm 0cm},clip]{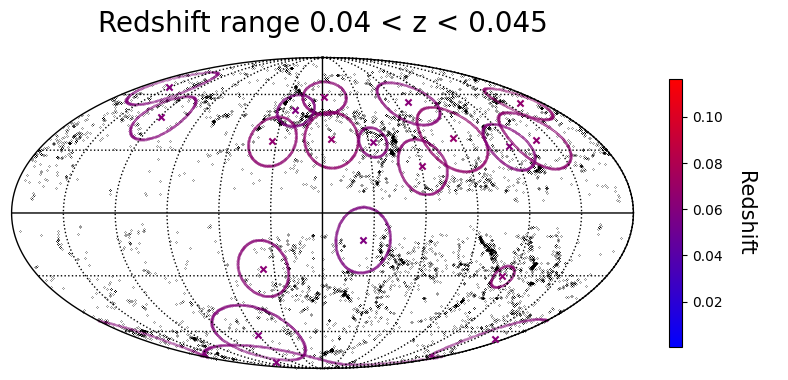}
        \\
        \vspace{0.5cm}
        \includegraphics[width=\linewidth, trim={0.2cm 0cm 3.7cm 0cm},clip]{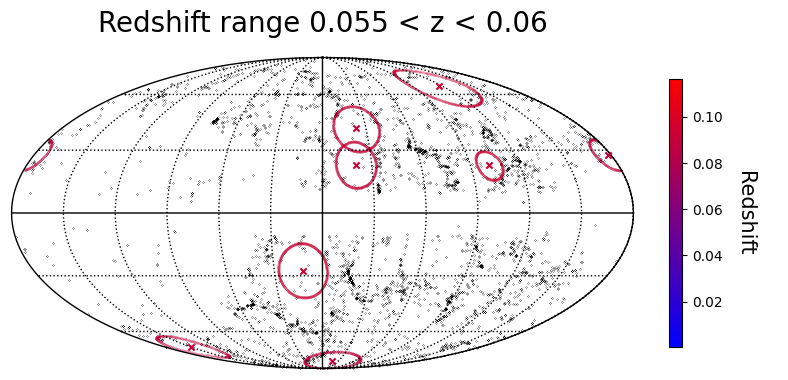}
        \\
        \vspace{0.5cm}
        \includegraphics[width=\linewidth, trim={0.2cm 0cm 3.7cm 0cm},clip]{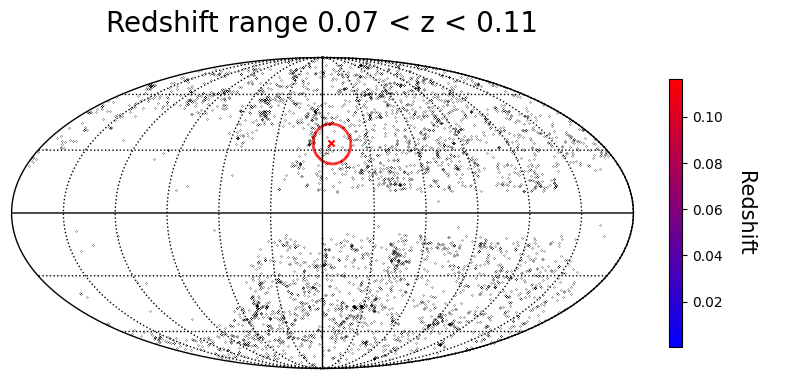}
        \\
        \vspace{0.5cm}
        
    \end{minipage}
    \begin{minipage}[t]{0.33\linewidth}
        \flushright
        \vspace{1cm}
        \includegraphics[width=\linewidth, trim={0.2cm 0cm 3.7cm 0cm},clip]{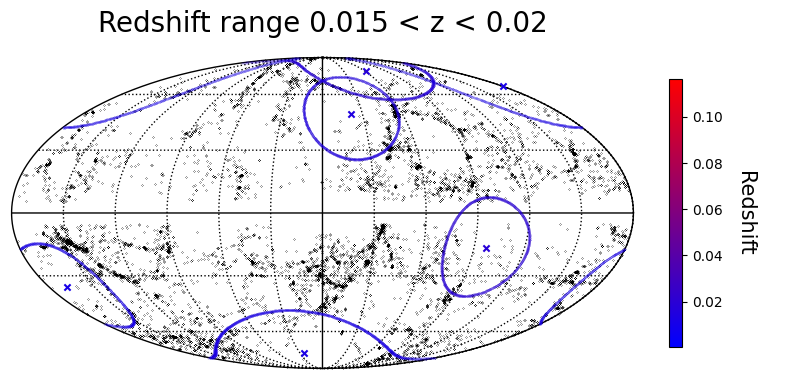}
        \\
        \vspace{0.5cm}
        \includegraphics[width=\linewidth, trim={0.2cm 0cm 3.7cm 0cm},clip]{figures/projections/zbin_5.png}
        \\
        \vspace{0.5cm}
        \includegraphics[width=\linewidth, trim={0.2cm 0cm 3.7cm 0cm},clip]{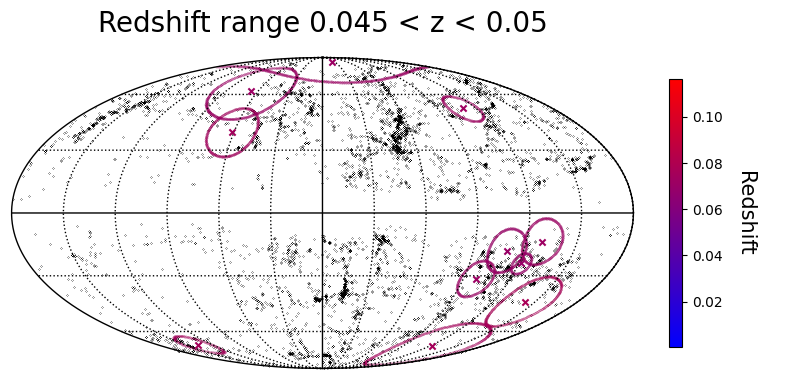}
        \\
        \vspace{0.5cm}
        \includegraphics[width=\linewidth, trim={0.2cm 0cm 3.7cm 0cm},clip]{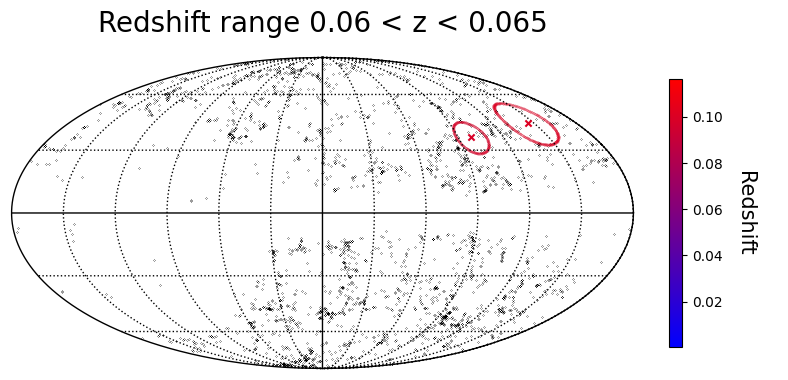}
        \\
        \vspace{0.5cm}
        
    \end{minipage}
    
    \caption{\\
    Different redshift slices representing the sky in galactic coordinates, with the zero longitude point in the center. The crosses represent the void centers, while the circles are the apparent size of their effective radii. The black dots represent the 2M++ galaxies used to infer the posterior realizations of the Local Universe. As a visual confirmation, voids actually occupy the regions with little to no galaxies. The predictive power of the \texttt{Manticore-Local} suite combined with our void finding procedure is shown by the fact that we are able to infer voids behind the galactic plane, in regions where we do not have direct observations of galaxies.
              }

    \label{fig:sky_projection_all}
\end{figure*}

In this section, we expand the discussion presented in the main text to the full catalog. Figure~\ref{fig:sky_projection_all} shows slices of the sky in equatorial coordinates in different redshift bins.
We also present the extended version of Table~\ref{tab:voids_catalog_short}, illustrating all the voids in the catalog, split in two halves in Table~\ref{tab:voids_catalog}~and~\ref{tab:voids_catalog_2}.
The first 3 columns represent the coordinates of the center in the sky, in terms of equatorial coordinates and distance.
Column 4 represents the redshift range from the nearest to the farther edge of the void along the line of sight, assuming the effective radius as isotropic size.
Columns 5-7 represent summary properties of the posterior distributions for the effective radius and the center position, using the mean and standard deviation.
In our cartesian coordinates system the observer is located at $[340.5, 340.5, 340.5]$ \Mpch, with the $xy$ plane corresponding to the equatorial plane, and the $\hat{z}$ axis pointing to the equatorial North Pole.
Visual representations of each void in terms of Voronoi clouds can be found in the last columns of the same table presented in the public website: \url{https://voids.cosmictwin.org/VoidGallery}.

\begin{table*}
\caption{First $50$ voids of the catalog, as described in Appendix~\ref{app:catalog}.}            
\label{tab:voids_catalog}      
\centering          
\begin{tabular}{c | c r c | c | c | c c c}     
\hline\hline 
\ & \multicolumn{3}{c | }{equatorial sky coordinates} & \multicolumn{1}{c | }{redshift }  & \multicolumn{1}{c | }{radius $r$} & \multicolumn{3}{c}{position $\vec{x}$}\\ 
\hline

ID & $ \alpha \, [\mathrm{hms}]$ &  $\delta \, [^\circ]$ &  $ d_c \, [\Mpch ]$ & $z$ range $[ \text{km s}^{-1} ]$ &  $\bar{r} \, [ \Mpch ]$ & $\bar{x} \, [ \Mpch ]$ & $\bar{y} \, [\Mpch ]$ & $\bar{z} \, [ \Mpch ]$ \\

\hline  
    $ 0 $ & $ 10^\text{h}12^\text{m}$ & $ -39.0$ & $ 159.6$ & $ 12 \, 800 - 19 \, 500 $ & $32.8 \pm 3.4$ & $229.9 \pm 7.0$ & $396.8 \pm 5.8$ & $240.2 \pm 5.1$ \\
    $ 1 $ & $ 23^\text{h}11^\text{m}$ & $ -12.4$ & $ 122.6$ & $ 8 \, 600 - 16 \, 200 $ & $37.7 \pm 5.8$ & $457.6 \pm 14.1$ & $315.3 \pm 9.7$ & $314.2 \pm 7.5$ \\
    $ 2 $ & $ 18^\text{h}36^\text{m}$ & $ 30.4$ & $ 105.9$ & $ 6 \, 900 - 14 \, 500 $ & $37.7 \pm 3.7$ & $354.9 \pm 5.6$ & $250.3 \pm 5.3$ & $394.0 \pm 7.0$ \\
    $ 3 $ & $ 10^\text{h}28^\text{m}$ & $ 25.3$ & $ 92.0$ & $ 6 \, 900 - 11 \, 600 $ & $23.0 \pm 3.2$ & $264.0 \pm 4.9$ & $373.1 \pm 3.1$ & $379.8 \pm 2.6$ \\
    $ 4 $ & $ 08^\text{h}40^\text{m}$ & $ 31.2$ & $ 107.0$ & $ 8 \, 100 - 13 \, 500 $ & $26.3 \pm 3.6$ & $281.5 \pm 4.4$ & $410.5 \pm 7.1$ & $395.9 \pm 6.7$ \\
    $ 5 $ & $ 11^\text{h}12^\text{m}$ & $ -21.1$ & $ 133.5$ & $ 9 \, 600 - 17 \, 400 $ & $38.5 \pm 4.3$ & $218.7 \pm 5.7$ & $366.2 \pm 9.0$ & $292.4 \pm 9.7$ \\
    $ 6 $ & $ 03^\text{h}50^\text{m}$ & $ 6.2$ & $ 83.4$ & $ 6 \, 900 - 9 \, 900 $ & $15.1 \pm 2.6$ & $385.0 \pm 3.2$ & $410.5 \pm 6.0$ & $349.5 \pm 3.1$ \\
    $ 7 $ & $ 17^\text{h}39^\text{m}$ & $ -55.6$ & $ 118.9$ & $ 8 \, 700 - 15 \, 300 $ & $32.6 \pm 4.0$ & $334.4 \pm 8.3$ & $273.6 \pm 6.2$ & $242.4 \pm 5.7$ \\
    $ 8 $ & $ 06^\text{h}55^\text{m}$ & $ -6.2$ & $ 44.2$ & $ 2 \, 000 - 6 \, 900 $ & $24.0 \pm 4.9$ & $330.0 \pm 4.2$ & $383.2 \pm 7.0$ & $335.7 \pm 6.9$ \\
    $ 9 $ & $ 13^\text{h}25^\text{m}$ & $ 69.6$ & $ 123.5$ & $ 9 \, 900 - 15 \, 100 $ & $25.4 \pm 3.6$ & $300.4 \pm 4.7$ & $324.9 \pm 5.3$ & $456.3 \pm 7.0$ \\
    $ 10 $ & $ 18^\text{h}05^\text{m}$ & $ -16.2$ & $ 38.8$ & $ 500 - 7 \, 300 $ & $33.9 \pm 3.0$ & $341.3 \pm 4.9$ & $303.2 \pm 3.4$ & $329.7 \pm 3.5$ \\
    $ 11 $ & $ 16^\text{h}50^\text{m}$ & $ 38.3$ & $ 143.8$ & $ 11 \, 400 - 17 \, 800 $ & $31.3 \pm 3.6$ & $306.6 \pm 4.5$ & $232.8 \pm 5.3$ & $429.6 \pm 7.5$ \\
    $ 12 $ & $ 02^\text{h}26^\text{m}$ & $ -15.0$ & $ 125.1$ & $ 9 \, 900 - 15 \, 400 $ & $26.7 \pm 3.0$ & $437.6 \pm 4.4$ & $412.4 \pm 4.5$ & $308.0 \pm 3.6$ \\
    $ 13 $ & $ 08^\text{h}58^\text{m}$ & $ 45.4$ & $ 62.7$ & $ 4 \, 300 - 8 \, 300 $ & $19.6 \pm 2.8$ & $309.6 \pm 2.9$ & $372.0 \pm 4.0$ & $385.1 \pm 3.6$ \\
    $ 14 $ & $ 09^\text{h}22^\text{m}$ & $ -77.9$ & $ 157.0$ & $ 11 \, 200 - 20 \, 600 $ & $46.0 \pm 5.3$ & $315.1 \pm 6.4$ & $361.4 \pm 16.8$ & $187.0 \pm 5.2$ \\
    $ 15 $ & $ 23^\text{h}48^\text{m}$ & $ 16.6$ & $ 95.3$ & $ 7 \, 500 - 11 \, 700 $ & $20.9 \pm 3.1$ & $431.8 \pm 4.7$ & $335.6 \pm 5.6$ & $367.7 \pm 2.9$ \\
    $ 16 $ & $ 04^\text{h}55^\text{m}$ & $ -2.8$ & $ 118.2$ & $ 9 \, 600 - 14 \, 300 $ & $23.4 \pm 2.7$ & $373.6 \pm 4.3$ & $453.8 \pm 5.1$ & $334.8 \pm 4.3$ \\
    $ 17 $ & $ 23^\text{h}35^\text{m}$ & $ -12.6$ & $ 78.6$ & $ 6 \, 500 - 9 \, 400 $ & $14.3 \pm 2.3$ & $416.8 \pm 1.6$ & $332.3 \pm 3.9$ & $323.4 \pm 4.0$ \\
    $ 18 $ & $ 23^\text{h}41^\text{m}$ & $ -28.3$ & $ 48.6$ & $ 2 \, 400 - 7 \, 300 $ & $24.3 \pm 4.0$ & $383.1 \pm 5.5$ & $337.0 \pm 3.6$ & $317.5 \pm 7.2$ \\
    $ 19 $ & $ 13^\text{h}06^\text{m}$ & $ 22.1$ & $ 138.6$ & $ 10 \, 000 - 18 \, 000 $ & $39.3 \pm 4.6$ & $217.3 \pm 7.7$ & $304.3 \pm 8.2$ & $392.6 \pm 4.7$ \\
    $ 20 $ & $ 05^\text{h}01^\text{m}$ & $ -9.1$ & $ 156.6$ & $ 13 \, 700 - 18 \, 000 $ & $21.1 \pm 3.1$ & $380.1 \pm 5.2$ & $489.9 \pm 4.3$ & $315.6 \pm 5.6$ \\
    $ 21 $ & $ 11^\text{h}16^\text{m}$ & $ 37.3$ & $ 157.6$ & $ 13 \, 000 - 18 \, 900 $ & $28.4 \pm 3.2$ & $217.4 \pm 6.9$ & $364.6 \pm 4.6$ & $435.9 \pm 4.1$ \\
    $ 22 $ & $ 14^\text{h}32^\text{m}$ & $ -48.3$ & $ 73.5$ & $ 5 \, 000 - 9 \, 800 $ & $24.0 \pm 2.2$ & $302.0 \pm 2.9$ & $310.4 \pm 2.9$ & $285.7 \pm 4.4$ \\
    $ 23 $ & $ 06^\text{h}49^\text{m}$ & $ -40.7$ & $ 136.4$ & $ 11 \, 300 - 16 \, 300 $ & $24.8 \pm 3.4$ & $318.6 \pm 5.7$ & $441.6 \pm 4.3$ & $251.7 \pm 4.0$ \\
    $ 24 $ & $ 21^\text{h}14^\text{m}$ & $ 76.3$ & $ 92.4$ & $ 5 \, 600 - 13 \, 100 $ & $36.8 \pm 3.6$ & $356.9 \pm 4.0$ & $326.0 \pm 6.5$ & $430.3 \pm 10.4$ \\
    $ 25 $ & $ 06^\text{h}31^\text{m}$ & $ -26.9$ & $ 97.4$ & $ 7 \, 600 - 12 \, 000 $ & $21.9 \pm 2.3$ & $328.9 \pm 3.5$ & $426.6 \pm 3.3$ & $296.3 \pm 4.1$ \\
    $ 26 $ & $ 01^\text{h}20^\text{m}$ & $ -47.9$ & $ 116.2$ & $ 8 \, 500 - 15 \, 000 $ & $32.2 \pm 3.9$ & $413.7 \pm 6.6$ & $367.2 \pm 4.2$ & $254.2 \pm 3.6$ \\
    $ 27 $ & $ 20^\text{h}07^\text{m}$ & $ 29.2$ & $ 59.9$ & $ 3 \, 500 - 8 \, 500 $ & $24.7 \pm 2.6$ & $368.0 \pm 4.2$ & $296.0 \pm 4.3$ & $369.7 \pm 4.2$ \\
    $ 28 $ & $ 10^\text{h}59^\text{m}$ & $ 2.8$ & $ 43.6$ & $ 2 \, 300 - 6 \, 400 $ & $20.2 \pm 2.0$ & $298.4 \pm 4.0$ & $351.9 \pm 4.2$ & $342.6 \pm 3.0$ \\
    $ 29 $ & $ 16^\text{h}44^\text{m}$ & $ 15.1$ & $ 130.5$ & $ 10 \, 300 - 16 \, 100 $ & $28.6 \pm 2.9$ & $299.7 \pm 4.5$ & $221.3 \pm 4.9$ & $374.6 \pm 5.3$ \\
    $ 30 $ & $ 12^\text{h}34^\text{m}$ & $ -40.5$ & $ 128.6$ & $ 9 \, 900 - 16 \, 100 $ & $30.0 \pm 4.2$ & $243.8 \pm 4.9$ & $326.1 \pm 5.2$ & $257.0 \pm 6.9$ \\
    $ 31 $ & $ 00^\text{h}37^\text{m}$ & $ -19.9$ & $ 122.6$ & $ 9 \, 400 - 15 \, 400 $ & $29.3 \pm 3.1$ & $454.3 \pm 5.6$ & $358.8 \pm 3.4$ & $298.7 \pm 6.3$ \\
    $ 32 $ & $ 20^\text{h}00^\text{m}$ & $ -64.2$ & $ 74.2$ & $ 5 \, 100 - 9 \, 800 $ & $23.5 \pm 3.7$ & $356.7 \pm 4.5$ & $312.5 \pm 5.0$ & $273.7 \pm 5.2$ \\
    $ 33 $ & $ 06^\text{h}15^\text{m}$ & $ 68.5$ & $ 90.7$ & $ 6 \, 100 - 12 \, 200 $ & $29.7 \pm 5.1$ & $338.3 \pm 4.1$ & $373.7 \pm 5.6$ & $424.9 \pm 9.1$ \\
    $ 34 $ & $ 09^\text{h}34^\text{m}$ & $ -6.0$ & $ 122.6$ & $ 9 \, 900 - 14 \, 900 $ & $24.8 \pm 3.1$ & $242.5 \pm 5.4$ & $413.0 \pm 3.7$ & $327.6 \pm 5.3$ \\
    $ 35 $ & $ 12^\text{h}04^\text{m}$ & $ 40.6$ & $ 89.5$ & $ 6 \, 900 - 11 \, 100 $ & $20.6 \pm 1.8$ & $272.5 \pm 2.5$ & $339.2 \pm 3.0$ & $398.8 \pm 3.3$ \\
    $ 36 $ & $ 12^\text{h}20^\text{m}$ & $ -6.6$ & $ 125.2$ & $ 9 \, 900 - 15 \, 400 $ & $27.0 \pm 2.2$ & $216.6 \pm 4.8$ & $329.6 \pm 6.7$ & $326.1 \pm 4.8$ \\
    $ 37 $ & $ 15^\text{h}28^\text{m}$ & $ 13.9$ & $ 119.7$ & $ 10 \, 200 - 14 \, 000 $ & $18.6 \pm 2.0$ & $269.0 \pm 4.1$ & $248.9 \pm 2.9$ & $369.2 \pm 1.5$ \\
    $ 38 $ & $ 03^\text{h}25^\text{m}$ & $ -62.0$ & $ 29.6$ & $ 500 - 5 \, 500 $ & $25.0 \pm 2.1$ & $349.2 \pm 4.8$ & $351.3 \pm 3.6$ & $314.4 \pm 2.8$ \\
    $ 39 $ & $ 04^\text{h}14^\text{m}$ & $ -20.7$ & $ 147.9$ & $ 11 \, 500 - 18 \, 400 $ & $33.5 \pm 3.9$ & $402.1 \pm 5.8$ & $464.3 \pm 4.5$ & $288.2 \pm 6.1$ \\
    $ 40 $ & $ 11^\text{h}24^\text{m}$ & $ 12.8$ & $ 170.8$ & $ 13 \, 800 - 20 \, 900 $ & $34.7 \pm 3.2$ & $176.0 \pm 8.1$ & $366.6 \pm 4.4$ & $378.5 \pm 10.0$ \\
    $ 41 $ & $ 21^\text{h}33^\text{m}$ & $ -11.2$ & $ 199.4$ & $ 17 \, 500 - 23 \, 000 $ & $26.4 \pm 2.8$ & $497.3 \pm 5.8$ & $223.7 \pm 5.6$ & $301.6 \pm 5.2$ \\
    $ 42 $ & $ 07^\text{h}22^\text{m}$ & $ -51.4$ & $ 58.2$ & $ 3 \, 300 - 8 \, 400 $ & $25.6 \pm 2.3$ & $327.7 \pm 5.2$ & $374.5 \pm 5.9$ & $295.0 \pm 5.1$ \\
    $ 43 $ & $ 14^\text{h}04^\text{m}$ & $ 15.0$ & $ 99.8$ & $ 7 \, 900 - 12 \, 200 $ & $21.1 \pm 2.5$ & $257.9 \pm 2.5$ & $290.8 \pm 4.1$ & $366.3 \pm 4.3$ \\
    $ 44 $ & $ 00^\text{h}24^\text{m}$ & $ -33.9$ & $ 164.8$ & $ 14 \, 400 - 19 \, 000 $ & $22.2 \pm 2.7$ & $476.5 \pm 4.5$ & $355.0 \pm 3.8$ & $248.6 \pm 4.2$ \\
    $ 45 $ & $ 12^\text{h}48^\text{m}$ & $ 12.9$ & $ 56.1$ & $ 3 \, 700 - 7 \, 500 $ & $19.0 \pm 2.8$ & $287.0 \pm 4.1$ & $329.2 \pm 2.6$ & $353.0 \pm 3.1$ \\
    $ 46 $ & $ 10^\text{h}59^\text{m}$ & $ 19.6$ & $ 117.0$ & $ 10 \, 100 - 13 \, 600 $ & $17.2 \pm 2.9$ & $234.2 \pm 4.5$ & $369.6 \pm 4.2$ & $379.8 \pm 3.7$ \\
    $ 47 $ & $ 10^\text{h}37^\text{m}$ & $ -15.4$ & $ 180.0$ & $ 15 \, 700 - 20 \, 800 $ & $24.7 \pm 3.2$ & $178.1 \pm 5.3$ & $401.7 \pm 4.1$ & $292.7 \pm 5.4$ \\
    $ 48 $ & $ 08^\text{h}15^\text{m}$ & $ 21.3$ & $ 154.7$ & $ 13 \, 100 - 18 \, 200 $ & $25.2 \pm 3.9$ & $260.3 \pm 4.5$ & $460.2 \pm 4.7$ & $396.7 \pm 6.4$ \\
    $ 49 $ & $ 14^\text{h}35^\text{m}$ & $ 7.3$ & $ 128.0$ & $ 10 \, 700 - 15 \, 200 $ & $21.9 \pm 3.8$ & $241.4 \pm 5.5$ & $261.1 \pm 4.2$ & $356.7 \pm 4.4$ \\

\hline                  
\end{tabular}
\end{table*}

\begin{table*}
\caption{Last $50$ voids of the catalog, as described in Appendix~\ref{app:catalog}.}            
\label{tab:voids_catalog_2}      
\begin{tabular}{c | c r c | c | c | c c c}     
\hline\hline 
\ & \multicolumn{3}{c | }{equatorial sky coordinates} & \multicolumn{1}{c | }{redshift }  & \multicolumn{1}{c | }{radius $r$} & \multicolumn{3}{c}{position $\vec{x}$}\\ 
\hline

ID & $ \alpha \, [\mathrm{hms}]$ &  $\delta \, [^\circ]$ &  $ d_c \, [\Mpch ]$ & $z$ range $[ \text{km s}^{-1} ]$ &  $\bar{r} \, [ \Mpch ]$ & $\bar{x} \, [ \Mpch ]$ & $\bar{y} \, [\Mpch ]$ & $\bar{z} \, [ \Mpch ]$ \\

\hline  
    $ 50 $ & $ 03^\text{h}17^\text{m}$ & $ 14.1$ & $ 47.1$ & $ 2 \, 900 - 6 \, 600 $ & $18.4 \pm 2.0$ & $370.3 \pm 1.9$ & $375.2 \pm 3.7$ & $351.9 \pm 3.2$ \\
    $ 51 $ & $ 00^\text{h}27^\text{m}$ & $ 42.0$ & $ 80.5$ & $ 5 \, 300 - 11 \, 000 $ & $28.2 \pm 3.5$ & $399.9 \pm 3.6$ & $347.6 \pm 4.8$ & $394.3 \pm 5.5$ \\
    $ 52 $ & $ 20^\text{h}04^\text{m}$ & $ -29.3$ & $ 172.6$ & $ 13 \, 400 - 21 \, 600 $ & $39.8 \pm 4.1$ & $417.9 \pm 9.9$ & $211.4 \pm 8.5$ & $256.0 \pm 11.0$ \\
    $ 53 $ & $ 15^\text{h}36^\text{m}$ & $ 29.9$ & $ 118.1$ & $ 9 \, 300 - 14 \, 500 $ & $25.4 \pm 3.4$ & $280.4 \pm 5.8$ & $257.7 \pm 4.7$ & $399.4 \pm 3.8$ \\
    $ 54 $ & $ 20^\text{h}33^\text{m}$ & $ -8.5$ & $ 126.3$ & $ 9 \, 700 - 15 \, 900 $ & $30.4 \pm 3.0$ & $418.0 \pm 5.6$ & $242.6 \pm 7.0$ & $321.8 \pm 6.0$ \\
    $ 55 $ & $ 06^\text{h}02^\text{m}$ & $ -32.4$ & $ 136.1$ & $ 12 \, 500 - 15 \, 000 $ & $12.1 \pm 1.9$ & $339.6 \pm 2.6$ & $455.5 \pm 2.1$ & $267.6 \pm 3.5$ \\
    $ 56 $ & $ 23^\text{h}07^\text{m}$ & $ -48.9$ & $ 109.9$ & $ 9 \, 000 - 13 \, 200 $ & $21.0 \pm 2.8$ & $410.9 \pm 3.2$ & $323.9 \pm 3.7$ & $257.7 \pm 3.3$ \\
    $ 57 $ & $ 11^\text{h}01^\text{m}$ & $ -29.4$ & $ 65.8$ & $ 3 \, 800 - 9 \, 400 $ & $27.9 \pm 3.0$ & $285.1 \pm 4.0$ & $355.2 \pm 6.1$ & $308.2 \pm 2.6$ \\
    $ 58 $ & $ 14^\text{h}49^\text{m}$ & $ 3.3$ & $ 157.6$ & $ 14 \, 000 - 17 \, 900 $ & $19.3 \pm 3.0$ & $224.2 \pm 3.6$ & $234.5 \pm 3.1$ & $349.5 \pm 3.8$ \\
    $ 59 $ & $ 10^\text{h}11^\text{m}$ & $ 30.0$ & $ 131.9$ & $ 11 \, 200 - 15 \, 500 $ & $20.9 \pm 2.2$ & $238.9 \pm 3.7$ & $392.7 \pm 4.6$ & $406.5 \pm 5.1$ \\
    $ 60 $ & $ 09^\text{h}49^\text{m}$ & $ -10.1$ & $ 155.0$ & $ 14 \, 300 - 17 \, 100 $ & $13.9 \pm 3.0$ & $212.2 \pm 3.1$ & $423.1 \pm 3.1$ & $313.3 \pm 3.2$ \\
    $ 61 $ & $ 01^\text{h}10^\text{m}$ & $ -6.1$ & $ 145.3$ & $ 13 \, 200 - 16 \, 200 $ & $14.5 \pm 2.6$ & $478.2 \pm 3.4$ & $384.1 \pm 3.1$ & $324.9 \pm 4.1$ \\
    $ 62 $ & $ 14^\text{h}24^\text{m}$ & $ -7.9$ & $ 52.5$ & $ 3 \, 000 - 7 \, 500 $ & $22.2 \pm 3.6$ & $298.4 \pm 3.2$ & $310.0 \pm 4.4$ & $333.3 \pm 4.4$ \\
    $ 63 $ & $ 15^\text{h}45^\text{m}$ & $ -70.4$ & $ 154.3$ & $ 13 \, 200 - 18 \, 000 $ & $23.2 \pm 4.5$ & $311.6 \pm 7.4$ & $297.4 \pm 4.5$ & $195.1 \pm 6.2$ \\
    $ 64 $ & $ 09^\text{h}35^\text{m}$ & $ 17.3$ & $ 183.7$ & $ 15 \, 100 - 22 \, 200 $ & $34.5 \pm 3.8$ & $199.0 \pm 4.8$ & $444.1 \pm 6.8$ & $395.2 \pm 5.6$ \\
    $ 65 $ & $ 14^\text{h}34^\text{m}$ & $ -14.8$ & $ 169.5$ & $ 13 \, 700 - 20 \, 700 $ & $34.3 \pm 2.6$ & $212.4 \pm 7.2$ & $238.4 \pm 6.4$ & $297.2 \pm 8.1$ \\
    $ 66 $ & $ 08^\text{h}45^\text{m}$ & $ -7.1$ & $ 151.1$ & $ 13 \, 600 - 17 \, 000 $ & $17.0 \pm 2.5$ & $241.5 \pm 2.5$ & $453.2 \pm 4.1$ & $321.9 \pm 3.2$ \\
    $ 67 $ & $ 15^\text{h}33^\text{m}$ & $ -10.6$ & $ 132.5$ & $ 10 \, 000 - 16 \, 800 $ & $33.1 \pm 3.8$ & $262.4 \pm 5.2$ & $236.3 \pm 6.6$ & $316.2 \pm 6.3$ \\
    $ 68 $ & $ 22^\text{h}48^\text{m}$ & $ 37.2$ & $ 116.5$ & $ 8 \, 900 - 14 \, 600 $ & $27.6 \pm 4.4$ & $428.8 \pm 6.7$ & $311.9 \pm 4.4$ & $410.9 \pm 5.6$ \\
    $ 69 $ & $ 11^\text{h}40^\text{m}$ & $ 48.3$ & $ 122.2$ & $ 10 \, 100 - 14 \, 600 $ & $22.2 \pm 2.5$ & $259.6 \pm 4.6$ & $347.7 \pm 4.1$ & $431.8 \pm 5.0$ \\
    $ 70 $ & $ 09^\text{h}17^\text{m}$ & $ 59.8$ & $ 102.5$ & $ 8 \, 600 - 12 \, 000 $ & $16.7 \pm 3.4$ & $301.5 \pm 4.0$ & $374.1 \pm 4.8$ & $429.1 \pm 4.5$ \\
    $ 71 $ & $ 07^\text{h}52^\text{m}$ & $ 38.7$ & $ 166.6$ & $ 14 \, 600 - 19 \, 100 $ & $21.9 \pm 2.8$ & $279.4 \pm 4.6$ & $455.3 \pm 3.5$ & $444.6 \pm 4.7$ \\
    $ 72 $ & $ 10^\text{h}53^\text{m}$ & $ 1.8$ & $ 140.7$ & $ 12 \, 500 - 16 \, 000 $ & $17.4 \pm 2.0$ & $205.8 \pm 3.5$ & $380.9 \pm 3.9$ & $344.8 \pm 4.9$ \\
    $ 73 $ & $ 12^\text{h}26^\text{m}$ & $ -79.8$ & $ 89.9$ & $ 5 \, 000 - 13 \, 100 $ & $39.6 \pm 3.8$ & $324.7 \pm 7.8$ & $338.7 \pm 4.0$ & $252.1 \pm 4.5$ \\
    $ 74 $ & $ 06^\text{h}33^\text{m}$ & $ -21.5$ & $ 139.1$ & $ 11 \, 400 - 16 \, 800 $ & $26.6 \pm 2.7$ & $321.8 \pm 6.2$ & $468.6 \pm 4.0$ & $289.5 \pm 4.0$ \\
    $ 75 $ & $ 00^\text{h}52^\text{m}$ & $ 43.9$ & $ 33.0$ & $ 1 \, 500 - 5 \, 100 $ & $18.0 \pm 2.0$ & $363.7 \pm 3.6$ & $345.8 \pm 2.1$ & $363.4 \pm 3.7$ \\
    $ 76 $ & $ 08^\text{h}11^\text{m}$ & $ 50.0$ & $ 110.5$ & $ 9 \, 600 - 12 \, 700 $ & $14.8 \pm 2.2$ & $302.2 \pm 2.7$ & $400.3 \pm 3.4$ & $425.2 \pm 2.3$ \\
    $ 77 $ & $ 22^\text{h}41^\text{m}$ & $ 8.6$ & $ 90.6$ & $ 7 \, 600 - 10 \, 700 $ & $15.5 \pm 2.7$ & $424.8 \pm 3.2$ & $310.3 \pm 3.8$ & $354.0 \pm 4.0$ \\
    $ 78 $ & $ 15^\text{h}38^\text{m}$ & $ -12.2$ & $ 207.1$ & $ 17 \, 400 - 24 \, 700 $ & $35.6 \pm 2.4$ & $223.0 \pm 7.7$ & $175.7 \pm 8.4$ & $296.7 \pm 10.0$ \\
    $ 79 $ & $ 13^\text{h}20^\text{m}$ & $ -11.0$ & $ 91.7$ & $ 7 \, 000 - 11 \, 500 $ & $22.2 \pm 4.2$ & $255.9 \pm 3.7$ & $309.6 \pm 4.4$ & $323.0 \pm 4.9$ \\
    $ 80 $ & $ 00^\text{h}32^\text{m}$ & $ -12.3$ & $ 193.8$ & $ 16 \, 400 - 23 \, 000 $ & $32.1 \pm 3.3$ & $528.0 \pm 6.8$ & $366.7 \pm 6.9$ & $299.3 \pm 7.0$ \\
    $ 81 $ & $ 14^\text{h}20^\text{m}$ & $ -24.5$ & $ 123.2$ & $ 10 \, 800 - 14 \, 100 $ & $16.2 \pm 2.0$ & $248.6 \pm 2.7$ & $276.2 \pm 3.8$ & $289.5 \pm 4.1$ \\
    $ 82 $ & $ 02^\text{h}21^\text{m}$ & $ -33.5$ & $ 145.1$ & $ 10 \, 900 - 18 \, 500 $ & $37.1 \pm 3.1$ & $439.2 \pm 6.2$ & $410.5 \pm 7.7$ & $260.3 \pm 7.7$ \\
    $ 83 $ & $ 17^\text{h}26^\text{m}$ & $ -10.9$ & $ 96.0$ & $ 8 \, 300 - 11 \, 100 $ & $14.0 \pm 2.6$ & $326.7 \pm 2.6$ & $247.2 \pm 2.3$ & $322.4 \pm 2.5$ \\
    $ 84 $ & $ 09^\text{h}51^\text{m}$ & $ -24.3$ & $ 173.1$ & $ 15 \, 400 - 19 \, 700 $ & $20.9 \pm 3.0$ & $207.1 \pm 4.0$ & $424.8 \pm 2.9$ & $269.2 \pm 4.3$ \\
    $ 85 $ & $ 15^\text{h}27^\text{m}$ & $ -28.6$ & $ 169.7$ & $ 13 \, 800 - 20 \, 600 $ & $32.7 \pm 2.8$ & $248.2 \pm 8.8$ & $223.6 \pm 4.3$ & $259.3 \pm 6.1$ \\
    $ 86 $ & $ 05^\text{h}40^\text{m}$ & $ 44.0$ & $ 44.7$ & $ 2 \, 500 - 6 \, 500 $ & $19.5 \pm 2.2$ & $343.2 \pm 2.5$ & $372.5 \pm 3.6$ & $371.6 \pm 3.1$ \\
    $ 87 $ & $ 03^\text{h}36^\text{m}$ & $ 53.8$ & $ 65.7$ & $ 4 \, 800 - 8 \, 400 $ & $18.1 \pm 2.2$ & $363.3 \pm 1.9$ & $371.9 \pm 2.2$ & $393.6 \pm 3.4$ \\
    $ 88 $ & $ 14^\text{h}51^\text{m}$ & $ 39.4$ & $ 138.6$ & $ 10 \, 000 - 18 \, 100 $ & $39.5 \pm 2.9$ & $261.9 \pm 10.7$ & $267.7 \pm 10.0$ & $428.5 \pm 9.3$ \\
    $ 89 $ & $ 11^\text{h}26^\text{m}$ & $ 12.0$ & $ 81.8$ & $ 6 \, 500 - 10 \, 000 $ & $17.5 \pm 2.6$ & $261.3 \pm 4.4$ & $352.3 \pm 4.1$ & $357.6 \pm 4.1$ \\
    $ 90 $ & $ 19^\text{h}12^\text{m}$ & $ -38.0$ & $ 106.4$ & $ 7 \, 000 - 14 \, 500 $ & $36.6 \pm 3.1$ & $366.4 \pm 7.8$ & $260.8 \pm 5.0$ & $275.0 \pm 5.4$ \\
    $ 91 $ & $ 12^\text{h}51^\text{m}$ & $ 20.9$ & $ 85.1$ & $ 6 \, 700 - 10 \, 400 $ & $18.1 \pm 1.9$ & $263.0 \pm 4.1$ & $323.0 \pm 3.6$ & $370.9 \pm 2.7$ \\
    $ 92 $ & $ 01^\text{h}28^\text{m}$ & $ -9.2$ & $ 167.3$ & $ 14 \, 700 - 19 \, 200 $ & $21.5 \pm 2.6$ & $493.6 \pm 5.5$ & $402.3 \pm 4.8$ & $313.9 \pm 3.7$ \\
    $ 93 $ & $ 09^\text{h}10^\text{m}$ & $ 10.4$ & $ 129.5$ & $ 9 \, 800 - 16 \, 400 $ & $32.6 \pm 3.0$ & $246.7 \pm 5.8$ & $426.7 \pm 5.7$ & $363.9 \pm 6.8$ \\
    $ 94 $ & $ 11^\text{h}42^\text{m}$ & $ 64.7$ & $ 83.3$ & $ 6 \, 300 - 10 \, 500 $ & $20.4 \pm 2.0$ & $305.0 \pm 4.4$ & $343.2 \pm 4.9$ & $415.8 \pm 4.8$ \\
    $ 95 $ & $ 05^\text{h}43^\text{m}$ & $ -54.2$ & $ 138.8$ & $ 11 \, 800 - 16 \, 200 $ & $21.4 \pm 1.9$ & $346.5 \pm 4.4$ & $421.4 \pm 2.0$ & $227.9 \pm 5.1$ \\
    $ 96 $ & $ 11^\text{h}05^\text{m}$ & $ 38.1$ & $ 46.6$ & $ 2 \, 600 - 6 \, 700 $ & $20.2 \pm 1.8$ & $304.9 \pm 4.6$ & $349.3 \pm 4.3$ & $369.3 \pm 4.5$ \\
    $ 97 $ & $ 12^\text{h}09^\text{m}$ & $ -10.0$ & $ 71.5$ & $ 5 \, 700 - 8 \, 600 $ & $14.3 \pm 1.3$ & $270.1 \pm 2.5$ & $337.8 \pm 3.4$ & $328.1 \pm 4.5$ \\
    $ 98 $ & $ 10^\text{h}24^\text{m}$ & $ 59.2$ & $ 112.3$ & $ 9 \, 200 - 13 \, 400 $ & $20.6 \pm 2.3$ & $288.0 \pm 5.0$ & $364.0 \pm 2.5$ & $436.9 \pm 3.9$ \\
    $ 99 $ & $ 05^\text{h}40^\text{m}$ & $ -40.4$ & $ 128.1$ & $ 11 \, 700 - 14 \, 200 $ & $12.2 \pm 2.8$ & $349.2 \pm 3.6$ & $437.7 \pm 4.5$ & $257.5 \pm 2.5$ \\
    
\hline                  
\end{tabular}
\end{table*}

\end{appendix}

\end{document}